%% file: main.tex
\documentclass[a4paper,onecolumn,11pt,unpublished,accepted=2023-02-07]{quantumarticle}
\pdfoutput=1
\usepackage{changes}
\usepackage{xspace}
\usepackage{qcircuit}
\usepackage{pythonhighlight}
\usepackage{env}

\usepackage{etoolbox}
\let\bbordermatrix\bordermatrix
\patchcmd{\bbordermatrix}{8.75}{4.75}{}{}
\patchcmd{\bbordermatrix}{\left ( }{\left[}{}{}
\patchcmd{\bbordermatrix}{\right ) }{\right]}{}{}

\newcommand{\perceval}{\textit{Perceval}\xspace}

\allowdisplaybreaks

\begin{document}

\title{Perceval: A Software Platform for Discrete Variable Photonic Quantum Computing}

%
    \author{Nicolas Heurtel}
   \email{nicolas.heurtel@quandela.com}
    \affiliation{Quandela, 7 Rue Léonard de Vinci, 91300 Massy, France}
    \affiliation{Université Paris-Saclay, Inria, CNRS, ENS Paris-Saclay,
    CentraleSupélec, LMF, 91190, 15 Gif-sur-Yvette, France}
%
    \author{Andreas Fyrillas}
    \affiliation{Quandela, 7 Rue Léonard de Vinci, 91300 Massy, France}
    \affiliation{Centre for Nanosciences and Nanotechnology, CNRS,
    Université Paris-Saclay, UMR 9001, 10 Boulevard Thomas Gobert,
    91120, Palaiseau, France}
    \author{Grégoire de Gliniasty}
    \affiliation{Quandela, 7 Rue Léonard de Vinci, 91300 Massy, France}
%
    \author{Raphaël Le Bihan}
    \affiliation{Quandela, 7 Rue Léonard de Vinci, 91300 Massy, France}
    \author{Sébastien Malherbe}
    \affiliation{Département de Physique de l’Ecole Normale Supérieure
    - PSL, 45 rue d’Ulm, 75230, Paris Cedex 05, France}
    \author{Marceau Pailhas}
    \affiliation{Quandela, 7 Rue Léonard de Vinci, 91300 Massy, France}

    \author{Eric Bertasi} 
    \affiliation{Quandela, 7 Rue Léonard de Vinci, 91300 Massy, France}
%
    \author{Boris Bourdoncle}
    \affiliation{Quandela, 7 Rue Léonard de Vinci, 91300 Massy, France}
    \author{Pierre-Emmanuel Emeriau}
    \affiliation{Quandela, 7 Rue Léonard de Vinci, 91300 Massy, France}
    \author{Rawad Mezher}
    \affiliation{Quandela, 7 Rue Léonard de Vinci, 91300 Massy, France}
    \author{Luka Music}
    \affiliation{Quandela, 7 Rue Léonard de Vinci, 91300 Massy, France}
%
    \author{Nadia Belabas}
    \affiliation{Centre for Nanosciences and Nanotechnology, CNRS,
    Université Paris-Saclay, UMR 9001, 10 Boulevard Thomas Gobert,
    91120, Palaiseau, France}
    \author{Benoît Valiron}
    \affiliation{Université Paris-Saclay, Inria, CNRS, ENS Paris-Saclay,
    CentraleSupélec, LMF, 91190, 15 Gif-sur-Yvette, France}
%
    \author{Pascale Senellart}
    \affiliation{Centre for Nanosciences and Nanotechnology, CNRS,
    Université Paris-Saclay, UMR 9001, 10 Boulevard Thomas Gobert,
    91120, Palaiseau, France}
%
    \author{Shane Mansfield}
    \email{shane.mansfield@quandela.com}
    \affiliation{Quandela, 7 Rue Léonard de Vinci, 91300 Massy, France}
%
    \author{Jean Senellart}
    \email{jean.senellart@quandela.com}
    \affiliation{Quandela, 7 Rue Léonard de Vinci, 91300 Massy, France}

\maketitle

\begin{abstract}
\noindent We introduce \perceval, an open-source software platform for simulating and interfacing with discrete-variable photonic quantum computers, and describe its main features and components.
Its Python front-end allows photonic circuits to be composed from basic photonic building blocks like photon sources, beam splitters, phase-shifters and detectors.
A variety of computational back-ends are available and optimised for different use-cases.
These use state-of-the-art simulation techniques
covering both weak simulation, or sampling, and strong simulation.
We give examples of \perceval in action by reproducing a variety of photonic experiments and simulating photonic implementations of a range of quantum algorithms, from Grover's and Shor's to examples of quantum machine learning.
\perceval is intended to be a useful toolkit for experimentalists wishing to easily model, design, simulate, or optimise a discrete-variable photonic experiment, for theoreticians wishing to design algorithms and applications for discrete-variable photonic quantum computing platforms, and for application designers wishing to evaluate algorithms on available state-of-the-art photonic quantum computers.
\end{abstract}

\maketitle

    \input{s1_introduction}

    \input{s2_LOQC}

    \input{s3_Percevalindepth}

    \input{s4_research}

    \input{s5_conclusion}

\section*{Acknowledgements}
The authors wish to thank Mario Valdiva for invaluable technical support, Arno Ricou for code and discussions on variational algorithms, and Jeanne Bourgeois, William Howard, Rayen Mahjoub, and Bechara Nasr, whose internships nourished early developments on the way to this paper. Finally, the authors would like to thank N.\ Quesada and the referees for their very valuable feedback that has greatly improved the quality of the paper.


\bibliographystyle{linksen}
\bibliography{perceval}

\newpage
\appendix

\input{appendix.tex}

\end{document}

%% file: s1_introduction.tex
\section{Introduction}

Quantum computing has gained huge interest and momentum in recent decades because of its promise to deliver computational advantages and speedups compared to the classical computing paradigm. The essential idea is that information processing takes place on physical devices,
and if the components of those devices behave according to the laws of quantum rather than classical physics
then it opens the door to exploiting quantum effects to process information in radically different and potentially advantageous ways.

Quantum algorithms like Shor's factorisation algorithm \cite{shor_algorithms_1994}, which gives an exponential speedup over its best known classical counterpart, or Grover's search algorithm \cite{grover_fast_1996}, which gives a quadratic speedup over its classical counterparts, are often cited, and have captured the imagination and provided motivation for the rapid developments that have taken place in quantum technologies.
However, for such algorithms to have practical significance will require large-scale fault-tolerant quantum computers.
Yet the quantum devices that are available commercially or in research laboratories today are somewhat more limited and belong, for now at least, to the so-called noisy-intermediate scale quantum (NISQ) regime \cite{preskill_quantum_2018}.

Of course NISQ devices are a necessary step on the path to large-scale fault-tolerant quantum computers,
but in principle they could already enable quantum computational advantages \cite{Preskill2011Quantum} -- in which they could outperform even the most powerful classical supercomputers available today -- for tasks with practical relevance.
Several experiments have already claimed to demonstrate quantum computational advantage in sampling tasks \cite{arute_quantum_2019,zhong_quantum_2020,wu_strong_2021,zhong_phase-programmable_2021, MadsenLaudenbach+2022}.
The practical significance of these is still being explored \cite{nikolopoulos_decision_2016,nikolopoulos_cryptographic_2019,banchi_molecular_2020},
but meanwhile many other promising proposals for quantum algorithms that may deliver practical advantages in this regime are also being pursued.
Among others, these include the quantum variational eigensolver \cite{peruzzo_variational_2014}, universal function approximators like that of \cite{yee_fock_2021}
(both of which we will return to later in this paper), the quantum approximate optimisation algorithm \cite{farhi_quantum_2014}, as well as a host of other algorithms \cite{BCK+2022NISQ} with applications that range from chemistry \cite{CRO+2019Quantum, MEA+2020Quantum} and many-body physics \cite{JSK+2018Quantum, DHM+2020Towards} to combinatorial optimisation \cite{VGS+2020Applying, ZHB+2020Adaptive} and machine learning \cite{SBI+2020Measuring, HBC+2021Quantum}.

In fact, a number of technological routes to building quantum computers are being actively pursued, in which quantum information is encoded in very different kinds of physical systems.
These include matter-based approaches that rely on superconducting circuits, cold atoms, or trapped ions, but also light-based approaches in which photons are the basic information-carrying systems.
Among these, photons have a privileged status, as they are the natural and indeed only viable support for communicating quantum information,
which will eventually be required for networking quantum processors and devices.
As such they will necessarily be a part of any longer-term developments in quantum computational hardware and infrastructure.
More than this, photons provide viable routes to both NISQ \cite{knill_scheme_2001} and large-scale fault-tolerant quantum computing through measurement-based models \cite{kieling_percolation_2007,bartolucci_fusion-based_2021} in their own right.

\perceval is a complete and efficient software platform for the discrete variable (DV) model for photonic quantum computing.\footnotemark It especially uses Fock state descriptions of photons generated by sources,
evolving through linear optical networks -- composed for example of beam splitters, phase shifters, waveplates or other linear optical components --
and then being detected.
The familiar qubit and measurement-based models for quantum computing can be encoded within the DV model (see \eg \cite{knill_scheme_2001,kieling_percolation_2007}).
However, the DV model is also of significant interest in its own right -- not least because it has led to some of the first proposals \cite{aaronson_computational_2011} for quantum computational advantage to be demonstrated with NISQ devices.
This concerns a computational problem known as Boson Sampling, discussed in detail in Section~\ref{sec:bs},
which essentially consists of sampling from the output distribution of photons that interfere in a generic interferometer.
We will also demonstrate direct DV photonic approaches to quantum machine learning in Section \ref{sec:qml}.

\footnotetext{\perceval does not aim to treat the continuous variable (CV) model of photonic quantum computing,
which is concerned with infinite dimensional observables of the electromagnetic field,
and which is the natural realm of the \textit{Strawberry Fields} platform \cite{killoran_strawberry_2019}.}

\perceval's features can be useful for designing, optimising, simulating, and eventually transpiling DV linear optical circuits and executing them on cloud-based physical processors. Although \perceval provides bridges (see Section \ref{subsub:qiskit}) with other open-source quantum computing toolkits \cite{fingerhuth_open_2018} such as Qiskit \cite{Qiskit}, it further allows users to work at a level that is closer to the photonic hardware than these toolkits regular gate-based qubit quantum circuit model,
which is already the focus of a number of other software platforms. It is also more flexible and supports many more functionalities than existing packages for linear optical quantum systems \cite{GGM+2021QOptCraft}.
The finer-grained control of the physical hardware can be valuable both for the NISQ regime,
where it is crucial to achieve the maximum performance from the specific hardware resources available,
and for optimising the building block photonic modules in schemes for reaching the large-scale fault-tolerant regime.

\perceval is intended to be useful for experimentalists wishing to design photonic experiments,
including allowing for realistic modelling of noise and imperfections,
and for computer scientists and theoreticians seeking to develop algorithms and applications for photonic quantum computers.
It is an open source platform that is intended for community development via \href{https://github.com/Quandela/Perceval}{GitHub}, the project \href{https://perceval.quandela.net/forum}{forum website} and an updated \href{https://perceval.quandela.net/docs}{documentation}.

\perceval integrates several state-of-the-art algorithms for running simulations optimised with low-level single instruction, multiple data (SIMD) implementations, allowing users to push close to the limits of classical simulability with desktop computers. Extensions to the framework are also intended for high-performance computing (HPC) cluster deployment which can permit simulation to scale further. Since version 0.7, \perceval is able to run samples from real photonic chips through the \href{https://cloud.quandela.com/}{Cloud}, giving the user access in real-time to the exact hardware characteristics of the photonic source (brightness, purity, indistinguishability), the chip, and the detectors. Remote computers are also available if the user wants to perform classical simulation with greater computational power.\footnote{More details on \url{https://perceval.quandela.net/docs/notebooks/Remote\%20computing.html}}

The remainder of this white paper is structured as follows.
We provide some brief background on photonic quantum computing in Section \ref{sec:pqc},
before outlining the structure and key features of \perceval in Section \ref{sec:perceval},
and then go on to give a number of illustrative examples of \perceval in action in Section \ref{sec:action}. The code of the examples can be found in Appendix \ref{app:examples}. This paper is based on version 0.7.3 of \perceval.

%% file: s2_LOQC.tex
\section{Photonic Quantum Computing}
\label{sec:pqc}

Similar to the qubit quantum circuit model,
the DV photonic model can be presented as a gate-based model,
in which states are prepared,
transformed as they are acted upon by gates,
and measured.

We consider a number $m \in \N$ of spatial modes.
Physically these could correspond to waveguides in an integrated circuit, optical fibres,
or paths in free space. Photon sources prepare initial states in these spatial modes.
These are number states $\ket{n}$, where $n$ is the number of photons in the mode,
or superpositions of number states.
We use the shorthand notation $\ket{0,1}$ for a two-mode system with state $\ket{0}$ in the first and $\ket{1}$ in the second, etc.
Unless otherwise specified, photons are assumed to be indistinguishable.
Sometimes we will wish to keep track of the polarisation of the photons,
which can be achieved by further splitting each spatial mode into two
polarisation modes, \eg horizontal (denoted by $\ket{H}$) and vertical (denoted by $\ket{V}$), and recording the
(superpositions of) number states for each.
\perceval also allows the possibility of tracking other attributes of photons.

Transformations are performed on the states by evolving them through linear optical networks.
The simplest linear optical operations (gates) are:
phase shifters,
which act on a single spatial mode,
and beam splitters,
which act on pairs of spatial modes.
These operations preserve photon number and are best described as unitary
matrices that act on the creation operator of each mode (see \cite{KMN+2007Linear, KL2010Introduction} for a more detailed treatment).
A creation operator is defined by its action $\mathbf{a}^\dagger \ket{n} =
\sqrt{n+1} \ket{n+1}$.
The unitary associated with a phase shifter $P_\phi$ with phase parameter $\phi \in
[0,2\pi]$ is simply the scalar $e^{i\phi}$,
while in the {\tt Rx} convention the unitary matrix associated to a beam splitter
$B_{\theta}$ is given in Equation \ref{eq:bs}. We have included for clarity the basis on which the matrix acts in gray above and on the left.

\begin{equation}
\label{eq:bs}
\widehat{U}_{\text{BS}}(\theta) = 
\bbordermatrix{
                            & \textcolor{gray}{\ket{1,0}}             & \textcolor{gray}{\ket{0,1}}                 \cr
\textcolor{gray}{\bra{1,0}} & \cos{\left(\frac{\theta}{2} \right)}              & i \sin{\left(\frac{\theta}{2} \right)}   \cr
\textcolor{gray}{\bra{0,1}} & i \sin{\left(\frac{\theta}{2} \right)} & \cos{\left(\frac{\theta}{2} \right)}                  \cr
} \, ,
\end{equation}
where the parameter $\theta$ relates to the reflectivity and $\ket{1,0}$ denotes the state in which the photon travels in the first spatial mode. \perceval introduces all theoretically equivalent beam splitter matrix conventions as shown in Table \ref{fig:components}. The action of any linear optical circuit is thus given by the unitary matrix obtained by composing and multiplying the matrices associated with its elementary components.
Interestingly, it can be shown that conversely any unitary evolution can be decomposed
into a combination of beam splitters and phase shifters,
\eg in a `triangular' \cite{Reck94} or `square' \cite{Clements16} array.

\perceval also allows for swaps or permutations of spatial modes,
and can include operations like waveplates,
which act on polarisation modes of a single spatial mode and are described by the following unitary \cite{ChekhovaBanzer+2021}:
\begin{equation}
\widehat{U}_{\text{WP}}(\delta, \xi) = 
\bbordermatrix{
        & \textcolor{gray}{\ket{H}}  &  \textcolor{gray}{\ket{V}} \cr
 \textcolor{gray}{\bra{H}} & i \sin{\left(\delta \right)} \cos{\left(2 \xi \right)} + \cos{\left(\delta \right)} & i \sin{\left(\delta \right)} \sin{\left(2 \xi \right)} \cr
 \textcolor{gray}{\bra{V}} & i \sin{\left(\delta \right)} \sin{\left(2 \xi \right)} & - i \sin{\left(\delta \right)} \cos{\left(2 \xi \right)} + \cos{\left(\delta \right)} \cr
} \, .
\end{equation}

Here $\delta$ is a parameter proportional to the thickness of the waveplate and $\xi$ represents the angle of the waveplate's optical axis in the $\left\{\ket{H}, \ket{V}\right\}$ plane. Especially important is the case that $\delta=\pi/2$, known as a half-wave plate, which rotates linear polarisations in the $\left\{\ket{H}, \ket{V}\right\}$ plane. Quarter-wave plates ($\delta=\pi/4$) convert circular polarisations to li\-near ones (\eg $\ket{L} = (\ket{H} + i \ket{V})/\sqrt{2}$) and vice-versa.

Polarising beam splitters will convert a superposition of polarisation modes in a single spatial mode to the corresponding equal-polarisation superposition of two spatial modes, and vice versa, and so in this sense allow us to translate between polarisation and spatial modes. The unitary matrix associated to a polarising beam splitter acting on the tensor product of the spatial mode and the polarisation mode is

\begin{equation}
\widehat{U}_{\text{PBS}} = 
\bbordermatrix{ 
  & \textcolor{gray}{\ket{H,0}} & \textcolor{gray}{\ket{V,0}} & \textcolor{gray}{\ket{0,H}} & \textcolor{gray}{\ket{0,V}} \cr
\textcolor{gray}{\bra{H,0}} &  0 & 0 & 1 & 0 \cr
\textcolor{gray}{\bra{V,0}} &  0 & 1 & 0 & 0 \cr
\textcolor{gray}{\bra{0,H}} &  1 & 0 & 0 & 0 \cr
\textcolor{gray}{\bra{0,V}} &  0 & 0 & 0 & 1 \cr
},
\end{equation}
where $\ket{H,0}$ denotes the single-photon state in which the photon travels in the first spatial mode with a horizontal polarisation.

Similarly then, any unitary evolution on spatial and polarisation modes can be decomposed into an array of polarising beam splitters, beam splitters and phase shifters.

Finally measurement, or readout, is made by single photon detectors
(these may be partially or fully number resolving).
Both photon generation and measurement are non-linear operations.
In particular, measurements can be used to induce probabilistic or post-selected non-linearities in linear optical circuits.
This could be used to further implement feedforward, whereby measurement events condition parameters further along in a circuit.

The basic post-selection-, feedforward- and polarisation-free (without polarisation mo\-des or operations) fragment of the DV model described above is precisely the model considered by Aaronson and Arkhipov in
\cite{aaronson_computational_2011}.
For classical computers, simulating this simple model, straightforward though its description may be,
is known to be $\#P$-hard,
and up to complexity-theoretic assumptions the same is true for approximate classical simulation.
The reason for this essentially comes down to the fact that calculating the detection statistics requires evaluating the permanents of unitary matrices,
a problem which itself is known to be $\#$P-hard \cite{valiant1979complexity}.
We describe this in a little more detail when we look at Boson Sampling in Section \ref{sec:bs}.

Similarly the polarisation-free fragment was shown by Knill, Laflamme and Milburn to be quantum computationally universal \cite{knill_scheme_2001}.
In particular, it can be used to represent qubits and qubit logic gates.

Just as the bit -- any classical two level system -- is generally taken as the basic informational unit in classical computer science,
the qubit -- any two level quantum system -- is usually taken as the basic informational unit in quantum computer science.
Photons have many degrees of freedom and offer a rich variety of ways to encode qubits.
One of the most common approaches is to use the dual-rail path encoding of \cite{knill_scheme_2001}.
Each qubit is encoded by one photon which may be in superposition over two spatial modes,
which correspond to the qubit's computational basis states:
\begin{equation}
\ket{0}_\mathrm{qubit} := \ket{1,0} \, , \quad
\ket{1}_\mathrm{qubit} := \ket{0,1} \, .
\label{eq:dual_rail_encoding}
\end{equation}
Single-qubit unitary gates are particularly straightforward to implement in this encoding, requiring only a fully parametrisable beam splitter, as in Equation \ref{eq:bs}, and a phase shifter.
An example of an entangling two-qubit gate ($\CNOT$), which uses heralding, \ie post-selection over auxiliary modes, will be presented in Section \ref{sec:perceval}.

Another common qubit encoding is the polarisation encoding,
and indeed versions of the KLM scheme also exist for this encoding \cite{spedalieri_linear_2005}.
Here each qubit is encoded in the polarisation degree of freedom of a single photon.
This encoding will be used in the example of Section \ref{subsec:grover}.
In practice, encodings may be chosen based on the availability or ease of implementation of different linear optical elements in laboratory settings.
For instance path-encoding is at present more accessible than polarisation encoding in integrated optics,
where linear optical circuits are implemented `on-chip'.
Although we have discussed ways of encoding qubit quantum circuits into the DV model,
it should be reiterated that the DV model has interesting features in its own right,
and the examples of Boson Sampling in Section \ref{sec:bs} and quantum machine learning in Section \ref{sec:qml} give some illustrations of interesting applications that bypass qubit descriptions.

We have presented the idealised DV model,
but it is important to note that in realistic implementations various noise, imperfections, and errors will arise.
\perceval is also intended as a tool for designing, modelling, simulating and optimising realistic DV linear optical circuits and experiments,
and to incorporate realistic noise-models.
We briefly demonstrate how \perceval handles imperfect photon purity at source and distinguishability due to imperfect synchronisation in Section \ref{sec:hom}.
Other relevant factors are losses at all stages from source to detection,
imperfect or imperfectly characterised linear optical components,
cross-talk effects,
detector dark-counts, etc.
Future versions of \perceval will contain increasingly sophisticated and parametrised modelling of such factors.

%% file: s3_Percevalindepth.tex
\section{Presentation of Perceval}
\label{sec:perceval}

\subsection{Global Architecture}

\perceval is a linear optical circuit development framework whose design is based on the following core ideas:
\begin{itemize}
\item It is simple to use, both for theoretical and experimental physicists and for computer scientists.
\item It does not constrain the user to any framework-specific conventions that are theoretically equivalent
(for example it was noticed early on that many different conventions for beam splitters can be found in the literature).
\item It provides state-of-the-art optimised algorithms -- as benchmarked on specific use cases.
\item It provides -- when possible and appropriate -- access to symbolic calculations for finding analytical solutions.
\item It provides companion tools, such as a unitary matrix toolkit, and \LaTeX or HTML rendering of algorithms.
\item It incorporates realistic, parameterisable error and noise modelling.
\item It aims to provide a seamless transition from simulators to actual photonic processors (QPU). As such, most of the programmatic interfaces are designed for QPU control. In particular, fixing a specific QPU automatically hides methods giving access to properties (such as probability amplitudes) and features which would be unavailable on the actual hardware.
\end{itemize}

\perceval is a modular object-oriented Python code, with optimised functions written in C making use of SIMD vectorisation. In the following section, we give an overview of the main classes available to the user. A full documentation is maintained on
\href{https://github.com/Quandela/Perceval}{GitHub} and available online through the project \href{https://perceval.quandela.net/docs}{website}.

\subsection{Main Classes}

\subsubsection{States}

Information in a linear optical circuit is encoded in the state of photons in certain ``modes'' that are defined by the circuit designer. \emph{States} are implemented in \perceval by the following two classes:

\begin{itemize}
    \item {\tt BasicState} is used to describe Fock states of $n$ photons over $m$ modes.By default, photons are indistinguishable but each photon can be annotated, controlling its distinguishability.
An annotation  is a way to associate additional information to
each photon and can thus represent additional degrees of freedom — for instance,
polarisation is represented as a photon annotation;
    \item {\tt StateVector} extends {\tt BasicState} to represent superpositions of states.
\end{itemize}

\subsubsection{Circuits}

\noindent The {\tt Circuit} class provides a practical way to assemble a linear optical circuit from predefined \emph{elementary components}, other circuits, or unitary matrices. It can also compute the unitary matrix associated to a given circuit, or conversely decompose a unitary into a linear optical circuit consisting of user-defined components. A library of predefined elementary components is provided (see Table \ref{fig:components}). Those components can be added on the desired mode(s) on the right of a {\tt Circuit} with the \pyth{Circuit.add(modes, component)} method. An example of code using this class can be found in Code \ref{code:cnot}.

\renewcommand{\arraystretch}{1.4}

\begin{table}[htp]
\centering
\begin{adjustbox}{max width=\textwidth, center}
\begin{tabular}{|c|c|c|}
\hline
Name                                                               & Unitary Matrix & Representation \\ \hline
\begin{tabular}[c]{@{}c@{}}\quad\\ Beam Splitter\\ \quad\end{tabular}& 
\begin{tabular}{l}
  \hspace{-0.8cm}$R_X$ convention: 
 \\ $\BSRx$ 
\\ \hspace{-0.8cm}$R_Y$ convention: \\  $\BSRy$ 
\\ \hspace{-0.8cm}$H$ convention: 
\\ $\BSH$ 
\end{tabular}
& 

\begin{tabular}{c}
  \vspace{-0.1cm}\\\hbox{\includegraphics[height=0.9cm]{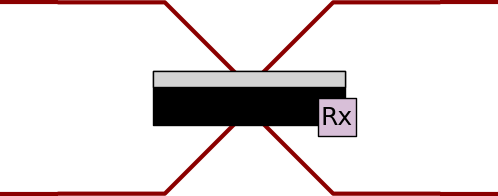}}  \\  
 \vspace{0.3cm}\\\hbox{\includegraphics[height=0.9cm]{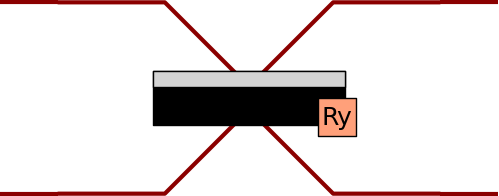}}\\ \vspace{0.3cm}\\\hbox{\includegraphics[height=0.9cm]{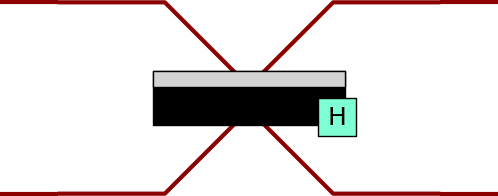}}
     
\end{tabular}
\\ \hline
Phase Shifter                                                      & $\begin{bmatrix}e^{i \phi}\end{bmatrix}$ & $\vcenter{\hbox{\includegraphics[height=0.7cm]{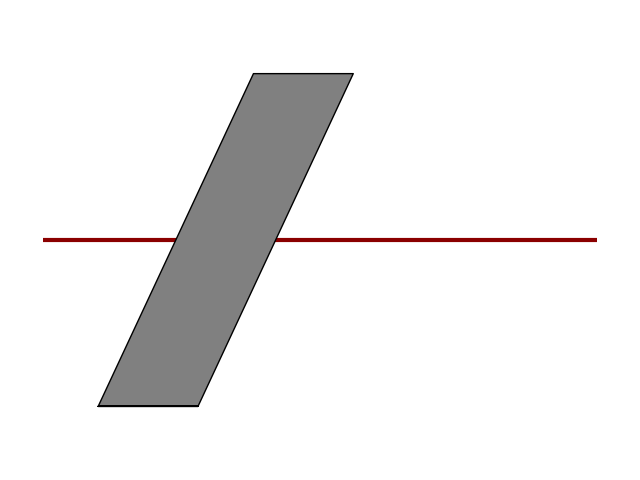}}}$\\ \hline
\begin{tabular}[c]{@{}c@{}}\quad\\ Mode Permutation\\ \quad\end{tabular}   & $\begin{bmatrix}0 & 1\\1 & 0\end{bmatrix}$ & $\vcenter{\hbox{\includegraphics[height=1cm]{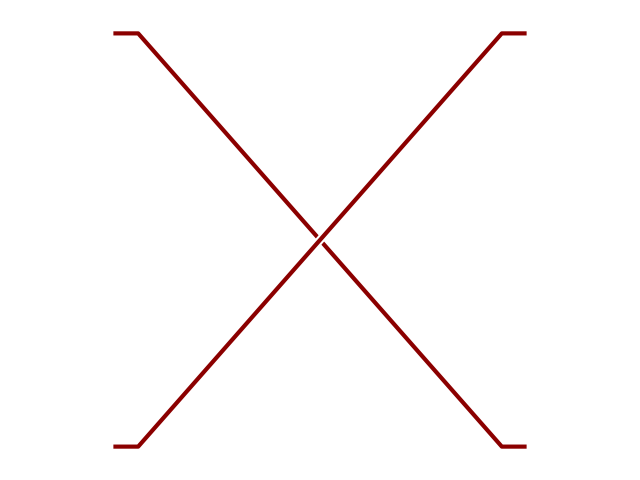}}}$\\ \hline
\begin{tabular}[c]{@{}c@{}}\quad\\ Wave Plate\\ \quad\end{tabular} & $\begin{bmatrix}i \sin{\left(\delta \right)} \cos{\left(2 \xi \right)} + \cos{\left(\delta \right)} & i \sin{\left(\delta \right)} \sin{\left(2 \xi \right)}\\i \sin{\left(\delta \right)} \sin{\left(2 \xi \right)} & - i \sin{\left(\delta \right)} \cos{\left(2 \xi \right)} + \cos{\left(\delta \right)}\end{bmatrix}$ & $\vcenter{\hbox{\includegraphics[height=0.8cm]{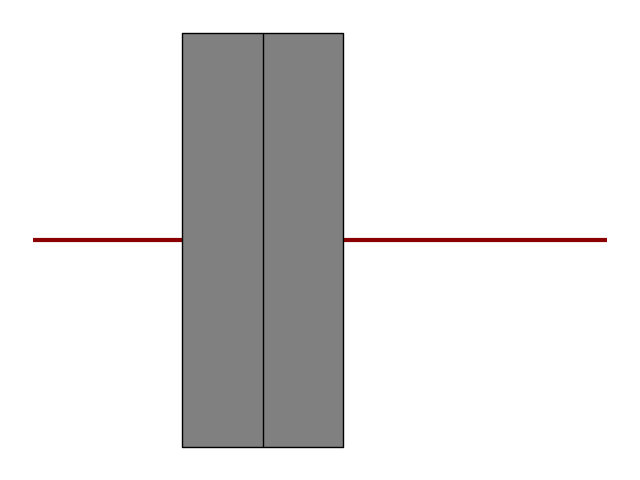}}}$\\ \hline
\begin{tabular}[c]{@{}c@{}}\quad\\ Polarising\\ Beam Splitter\\ \quad\end{tabular} & $\begin{bmatrix}0 & 0 & 1 & 0\\0 & 1 & 0 & 0\\1 & 0 & 0 & 0\\0 & 0 & 0 & 1\end{bmatrix}$ & $\vcenter{\hbox{\includegraphics[height=1.5cm]{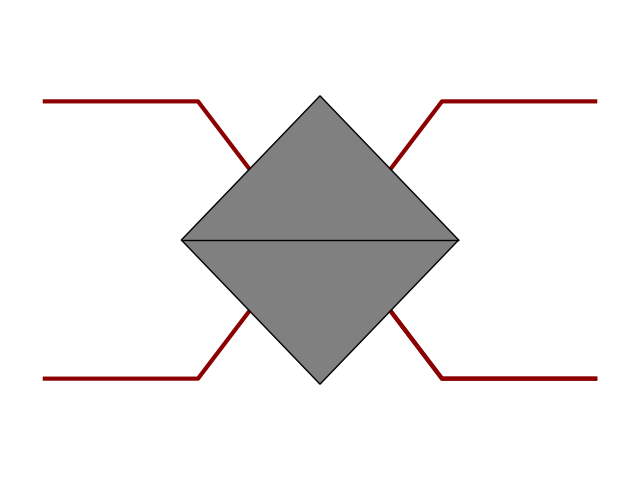}}}$\\ \hline
\begin{tabular}[c]{@{}c@{}}\quad\\ Polarising Rotator\\ \quad\end{tabular}       & $\begin{bmatrix}\cos{\left(\delta \right)} & \sin{\left(\delta \right)}\\- \sin{\left(\delta \right)} & \cos{\left(\delta \right)}\end{bmatrix}$ & $\vcenter{\hbox{\includegraphics[height=1cm]{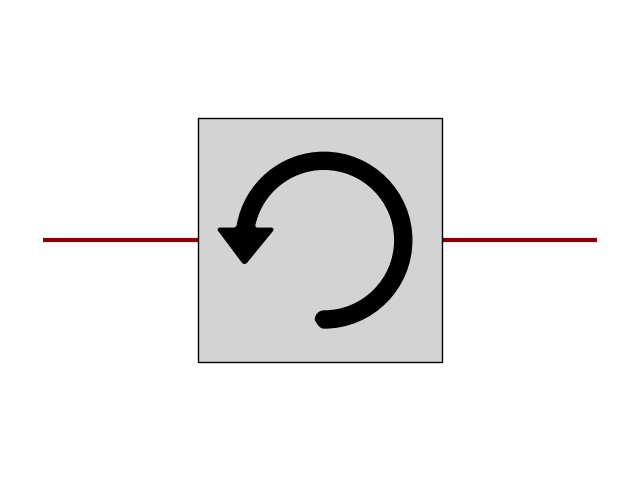}}}$\\ \hline
Time Delay                                                         &                & $\vcenter{\hbox{\includegraphics[height=0.7cm]{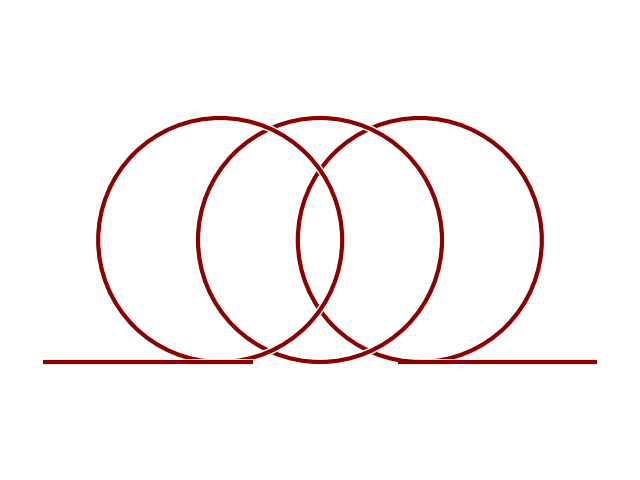}}}$\\ \hline
\end{tabular}
\end{adjustbox}
\caption{Components available in the {\tt components} library. Note that the beam splitter differs from the one given in Equation \ref{eq:bs}. The extra parameters allow the user to fix their own preferred convention for the beam splitter. \perceval includes a library which allows the user to define their own personal set of components, each with its own visual representation and unitary matrix. By default, all the phases $\phi$ of the beam splitters are set to zero.}
\label{fig:components}
\end{table}

\renewcommand{\arraystretch}{1}

\subsubsection{Back-ends}
\label{subsub:back}
\perceval allows the user to choose between its four \emph{back-ends} -- {\tt CliffordClifford2017}, {\tt Naive}, {\tt SLOS} and {\tt Stepper} -- each one taking a different computational approach to circuit simulation. They perform the following tasks:

\begin{itemize}
    \item Sample individual single output states -- {\tt CliffordClifford2017};
    \item Compute the probability, or probability amplitude, of obtaining a given output state from a given input state -- {\tt Naive};
    \item Describe the exact complete output state -- {\tt SLOS}, {\tt Stepper}.
\end{itemize}

\paragraph{The {\tt CliffordClifford2017} Back-end.}

This back-end is the implementation of the algorithm introduced in \cite{clifford_classical_2018}. The algorithm, applied to Boson Sampling, aims to ``\emph{produce provably correct random samples from a particular quantum mechanical distribution}''. Its time and space complexity are respectively $\bigO{n2^n+mn^2}$ and $\bigO{m}$. The algorithm has been implemented in C++, and uses an adapted Glynn algorithm \cite{glynn2010permanent} to efficiently compute $n$ simultaneous ``sub-permanents''.

Recently, the same authors have proposed a faster algorithm in \cite{clifford2020faster} with an average time complexity of $\bigO{n\rho_\theta^n}$ for a number of modes $m=\theta n$ which is linear in the number of photons $n$, where: 
\begin{equation}
    \rho_\theta = \frac{(2\theta+1)^{2\theta+1}}{(4\theta)^{\theta}(\theta+1)^{\theta+1}}
\end{equation}

For example, taking $\theta=2$, which corresponds to dual rail path encoding without auxiliary modes, the average performance of this algorithm is $\bigO{n(\frac{5^5}{8^23^3})^n} \approx \bigO{n1.8^n}$ as opposed to $\bigO{n2^n}$ for the original algorithm of \cite{clifford_classical_2018}. The implementation of \cite{clifford2020faster} in \perceval is in progress.

\paragraph{The {\tt Naive} Back-end.}

\begin{figure}
    \centering
    \includegraphics[width=10cm]{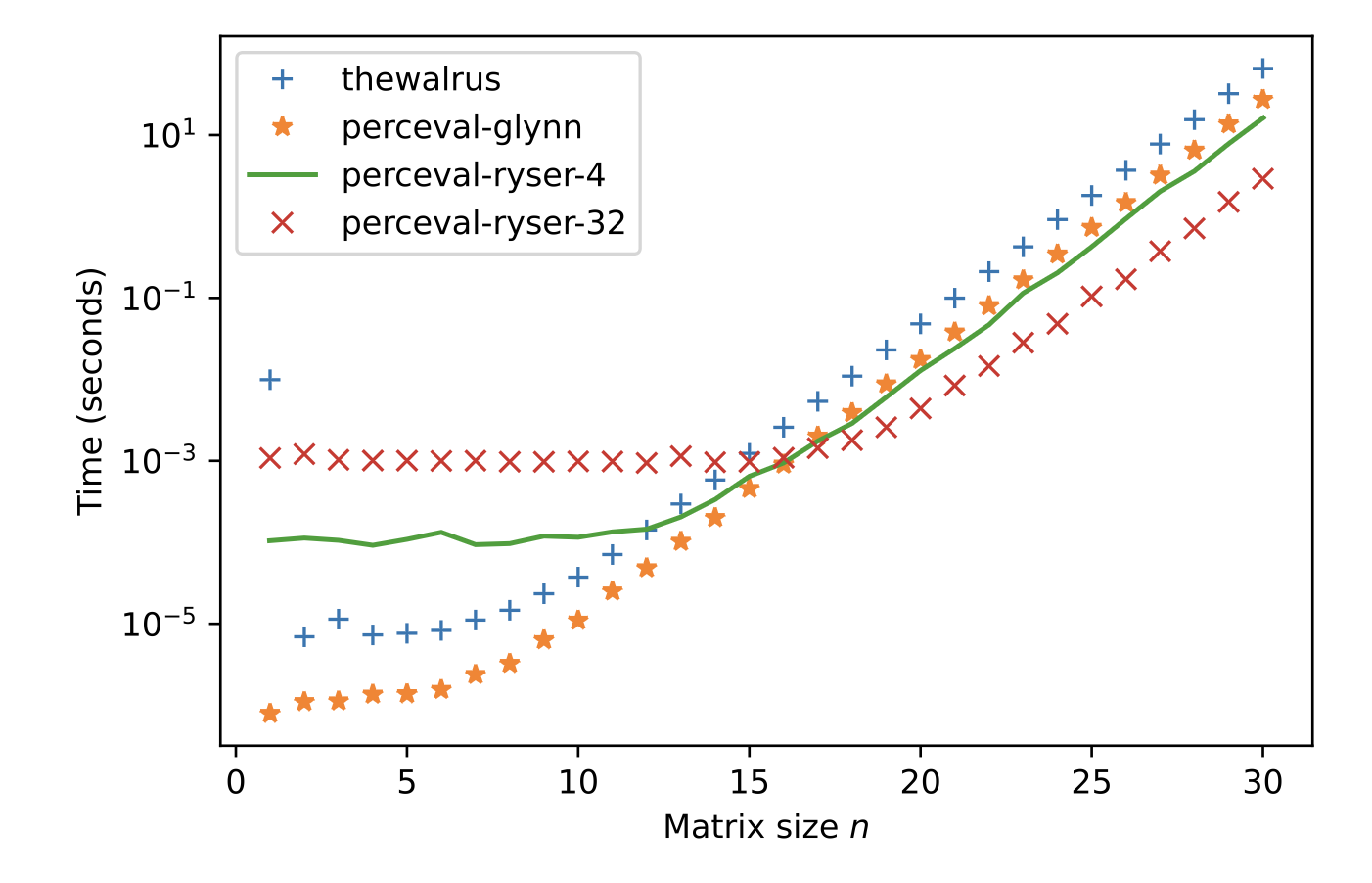}
    \caption{Comparison of the average time to calculate a permanent\protect\footnotemark of an $n\times n$ Haar random matrix. The processor is a 32 core, 3.1GHz Intel Haswell. For {\it The Walrus}, version 0.19 is used and installed from {\tt pypi}. The Ryser implementation is run on 4 or 32 threads. The Glynn implementation is run on a single thread. What is interesting to note is that all implementations have convergence to the theoretical performance but the factor between optimised and less optimised implementation still makes a perceptible time difference for the end-user. Based on different behaviour between Ryser and Glynn with $n$ and potential multi-threading, \perceval has some built-in logic to switch between the two algorithms.}
    \label{fig:performance-permanent}
\end{figure}

This back-end implements direct permanent calculation and is therefore suited for single output probability computation with small memory cost. Both Ryser's \cite{ryser1963combinatorial} and Glynn's \cite{glynn2010permanent} algorithms have been implemented. Extra care has been taken on the implementation of these algorithms, with usage of different optimisation techniques including native multithreading and SIMD vectorisation primitives. A benchmark of these algorithms against the implementation present in the \textit{The Walrus} \href{https://github.com/XanaduAI/thewalrus}{library} \cite{Gupt2019} is provided in Figure~\ref{fig:performance-permanent}.

\footnotetext{Following the methodology presented at \url{https://the-walrus.readthedocs.io/en/latest/gallery/permanent_tutorial.html}.}

\paragraph{The {\tt SLOS} Back-end.}

The Strong Linear Optical Simulation {\tt SLOS} algorithm developed by a subset of the present authors is introduced in \cite{heurtel2022}. It unfolds the full computation path in memory, leading to a remarkable time complexity of $\bigO{n\binom{m+n-1}{n}}$ for computing the full distribution. The current implementation also allows restrictive sets of outputs, with average computing time in $\bigO{n\rho_\theta^n}$ for single output computation. As discussed in \cite{heurtel2022}, it is possible to use the {\tt SLOS} algorithm in a hybrid manner that can combine both weak and strong simulation, though it has not yet been implemented in the current version of \perceval.
The tradeoff in the {\tt SLOS} algorithm is a huge memory usage of $\bigO{n\binom{m+n-1}{n}}$ that limits usage to circuits with $\approx 20$ photons on personal computers and with $\approx 24$ photons on super-computers.

\paragraph{The {\tt Stepper} Back-end.} This back-end takes a completely different approach. Without computing the circuit's overall unitary matrix first, it applies the unitary matrix associated with the components (see Table \ref{fig:components}) in each layer of the circuit one-by-one, simulating the evolution of the state vector. The complexity of this back-end is therefore proportional to the number of components. It has the nice features that:

\begin{itemize}
    \item it can support more complex components like Time Delay;
    \item it is very flexible with simulating noise in the circuit, like photon loss, or more generally with simulating any non-linear operation the user would wish to implement;
    \item it simplifies the debugging of circuits by exposing intermediate states.
\end{itemize}

The \texttt{Stepper} back-end is really meant for circuit simulation with losses or non linear components. In all other cases, it is more efficient to directly consider the unitary matrix of the whole circuit and compute the output state with {\tt SLOS}, instead of performing a computation for each component using 
 the \texttt{Stepper} back-end.

Theoretical performances and specific features of the different back-ends are summarised in Table \ref{fig:performance-table}.

\renewcommand{\arraystretch}{1.2}

\begin{table}[ht]
\centering
\begin{adjustbox}{max width=\textwidth, center}
\begin{tabular}{|c|c|c|c|c|}
\hline
Feature                                                                & \tt CC2017 & \tt SLOS & \tt Naive & \tt Stepper \\ \hline
\begin{tabular}[c]{@{}c@{}}Sampling\\ Efficiency\end{tabular}          & $\bigO{n2^n+p(m,n^2)} (*) $ & (**) & $\bigO{n2^n\binom{m+n-1}{n}}$  &                           \\ \hline
\begin{tabular}[c]{@{}c@{}}Single Output\\ Efficiency\end{tabular}     & N/A                  & (**) & $\bigO{n2^n}$ & $\smallO{N_c\binom{m+n-1}{n}}$ \\ \hline
\begin{tabular}[c]{@{}c@{}}Full Distribution\\ Efficiency\end{tabular} & N/A                  & $\bigO{n\binom{m+n-1}{n}}$ & $\bigO{n2^n\binom{m+n-1}{n}}$ & $\smallO{N_c\binom{m+n-1}{n}}$ \\ \hline
\begin{tabular}[c]{@{}c@{}}Probability\\ Amplitude\end{tabular}        & No         & Yes      & Yes       & Yes         \\ \hline
\begin{tabular}[c]{@{}c@{}}Supports Symbolic\\ Computation\end{tabular} & No         & No       & Yes       & Yes         \\ \hline
\begin{tabular}[c]{@{}c@{}}Supports\\ Time-Circuit\end{tabular}      & No         & No       & No        & Yes         \\ \hline
\begin{tabular}[c]{@{}c@{}}Practical\\ Limits (***)\end{tabular}   & $n \approx 30$       & $n, m < 20$           & $n \approx 30$         &                           \\ \hline
\end{tabular}
\end{adjustbox}
\caption{Theoretical performance and specific features of \perceval back-ends. (*) An implementation based on ``{\it Faster Classical Boson Sampling}'' \cite{clifford2020faster} is in progress; $p$ is a polynomial function. (**) An implementation of SLOS Sampling and single output efficiency is in progress. (***) Practical limits are subjective and corresponding to a memory usage $< 16Gb$, and a usage time for a given function of less than a few seconds. For {\tt Stepper}, it is hard to evaluate exactly the complexity since it is really proportional to the number of ``components'' ($N_c$) and the size of the output space that are circuit-specific.}
\label{fig:performance-table}
\end{table}

\renewcommand{\arraystretch}{1}

\subsubsection{Processors}
\label{subsub:processors}

The {\tt Processor} class allows the user to emulate a real photonic quantum processor, taking into account the single photon source, the circuit and the detectors. The {\tt Perceval} source model allows tuning of the transmittance, multiple photon emission probability, and indistinguishability. By default, the source is perfect. The circuit can be composed of unitary or non-unitary components. Detector imperfections can be emulated -- one can switch between number resolving detectors and threshold detectors. Post-selection features can also be added to a {\tt Processor} via heralded modes and/or a final post-selection function. See the Appendix \ref{app:proc} for more details.

\subsubsection{Algorithms}

The {\tt algorithm} library provides a set of tasks which can be performed on a {\tt Processor}.
These tasks can be as simple as obtaining a sample result (see Appendix \ref{app:proc}), or slightly more complex {\tt Analyzer} will output a probability table of expected input-output correlations, as illustrated in Figure \ref{fig:ralph_cnot_simul_result}.

\subsubsection{Bridges to Other Quantum Computing Toolkits}
\label{subsub:qiskit}

In order to facilitate work across multiple platforms, \perceval provides both a catalog of predefined components and specific converters:

\begin{itemize}
\item The predefined component catalog is available in {\tt Perceval.components.catalog}, where each element is an instance of {\tt CatalogItem} providing a ready-to-use circuit ({\tt as\_circuit()} method), a processor ({\tt as\_processor()} method), and a complete documentation ({\tt doc} property).
\item Converters to \perceval from other frameworks are available in {\tt perceval.converters}. Their task is to provide a photonic \perceval equivalent of circuits and algorithms defined in other open source frameworks. Currently a {\tt qiskit}\footnote{https://qiskit.org} converter is available and converters for \href{https://github.com/CQCL/tket}{\tt tket} and \href{https://myqlm.github.io}{\tt QLM} are in development.
\end{itemize}

Examples for both tools are provided in a \href{https://perceval.quandela.net/docs/notebooks/Qiskit\%20conversion.html}{notebook} of the \perceval \href{https://perceval.quandela.net/docs/}{documentation}.

\subsubsection{Step-by-Step Example}
\label{sec:stepbystep}

The following code implements a simple linear optical circuit corresponding to a path-encoded $\CNOT$ gate (after post-selection in the coincidence basis) \cite{ralph_linear_2002}. 

\begin{lstlisting}[language=python,style=mypython,caption= Implementation of the $\CNOT$ of \cite{ralph_linear_2002},captionpos=b,label=code:cnot]
import perceval as pcvl
from perceval.components import BS
import numpy as np

theta_13 = BS.r_to_theta(r=1/3)
cnot = (pcvl.Circuit(6, name="Ralph CNOT")
        .add((0, 1), BS.H(theta_13, phi_bl=np.pi,
                               phi_tr=np.pi/2, phi_tl=-np.pi/2))
        .add((3, 4), BS.H())
        .add((2, 3), BS.H(theta_13, phi_bl=np.pi,
                               phi_tr=np.pi/2, phi_tl=-np.pi/2))
        .add((4, 5), BS.H(theta_13))
        .add((3, 4), BS.H()))
pcvl.pdisplay(cnot) #displays the circuit 
pcvl.pdisplay(cnot.compute_unitary(), output_format=pcvl.Format.LATEX)
#outputs the unitary matrix of the circuit
\end{lstlisting}
    
\begin{figure}[ht]
    \centering
    \includegraphics[height=6cm]{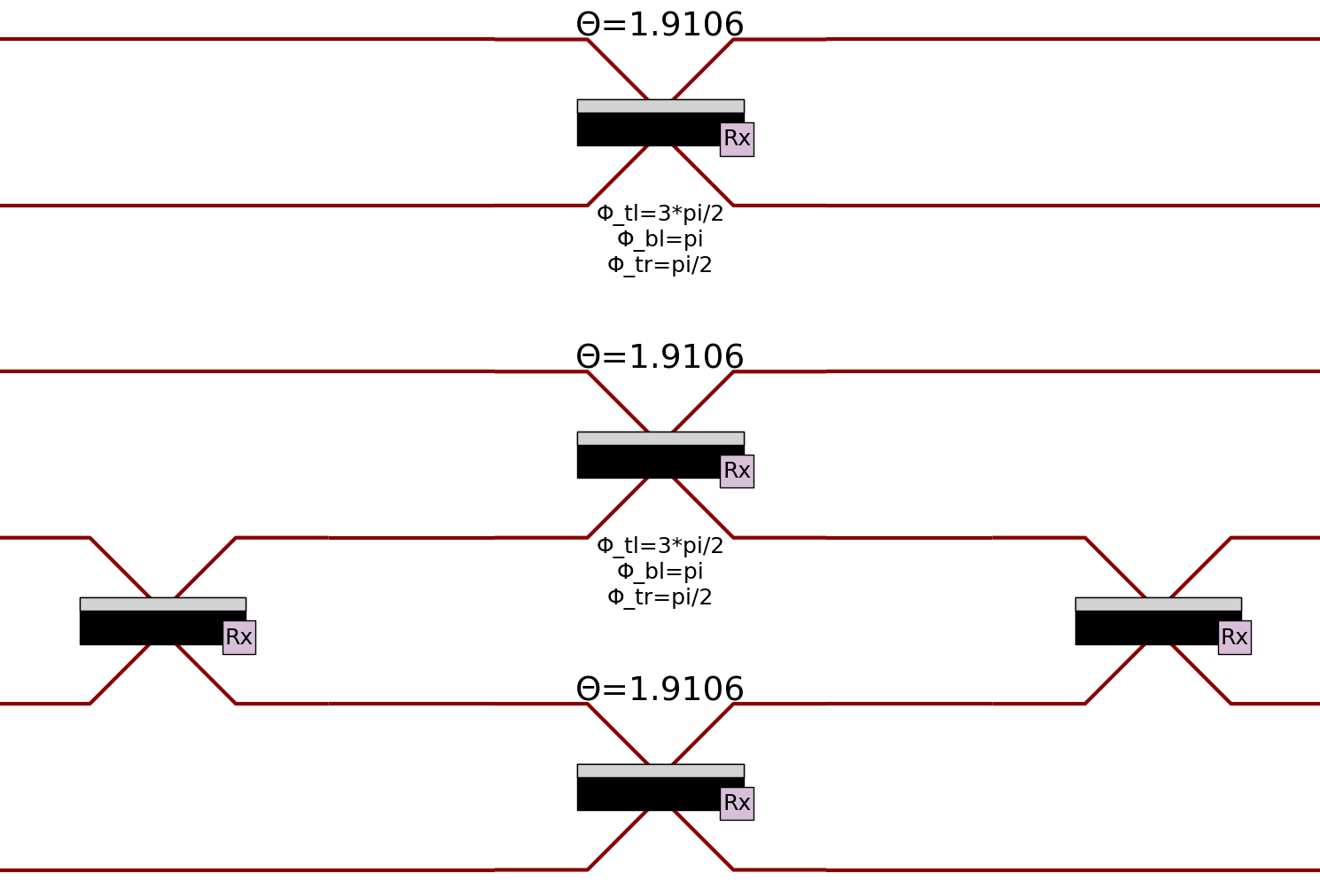}
    \caption{The circuit generated by \perceval, output of the Code \ref{code:cnot}.}
    \label{fig:ralph_cnot_circuit}
\end{figure}

Let us explain how it works. We remind that the $\CNOT$ gate is a two-qubit gate that acts in the following way: 

\begin{equation}
    \CNOT(\ket{x,y})=\ket{x,x\oplus y},
\end{equation}
\ie it's a controlled-not gate where the first qubit acts as control and the second qubit is the target. We use dual-rail encoding (see Equation \ref{eq:dual_rail_encoding}) and thus need four spatial modes to represent the target and the control. We additionally need two auxiliary empty modes to implement this post-selected linear optical $\CNOT$. The circuit is made of three central beam splitters of reflectivity $2/3$, whereas the two to the left and right of the central ones have a reflectivity  of $1/2$. The corresponding circuit displayed by \perceval is depicted in Figure \ref{fig:ralph_cnot_circuit}. 

The first and last spatial modes correspond to the auxiliary empty modes, the second and third to the control qubit, and the fourth and fifth to the target qubit. The associated unitary matrix acting on the six spatial modes computed with \perceval is displayed in Figure \ref{fig:ralph_cnot_unitary}. 

\begin{figure}[ht]
    \centering
    $$\left[\begin{matrix}\frac{\sqrt{3}}{3} & - \frac{\sqrt{6} i}{3} & 0 & 0 & 0 & 0\\- \frac{\sqrt{6} i}{3} & \frac{\sqrt{3}}{3} & 0 & 0 & 0 & 0\\0 & 0 & \frac{\sqrt{3}}{3} & - \frac{\sqrt{3} i}{3} & - \frac{\sqrt{3} i}{3} & 0\\0 & 0 & - \frac{\sqrt{3} i}{3} & \frac{\sqrt{3}}{3} & 0 & \frac{\sqrt{3}}{3}\\0 & 0 & - \frac{\sqrt{3} i}{3} & 0 & \frac{\sqrt{3}}{3} & - \frac{\sqrt{3}}{3}\\0 & 0 & 0 & \frac{\sqrt{3}}{3} & - \frac{\sqrt{3}}{3} & - \frac{\sqrt{3}}{3}\end{matrix}\right]$$
    \caption{The unitary matrix computed symbolically, output of the Code \ref{code:cnot}.} \label{fig:ralph_cnot_unitary}
\end{figure}

We then need to post-select on measuring two (dual-rail encoded) qubits in the modes corresponding to the target and the control, \ie on measuring exactly one photon in the second or third spatial modes and exactly one photon in the fourth or fifth spatial modes. One can check that this measurement event happens with probability $1/9$ -- we call this number the `performance' of a post-selected gate. We say that the error rate is $0$ if the implementation of the gate is perfect (after post-selection). This computation can be done with \perceval:  we define a processor, use the {\tt SLOS} back-end, and perform a full output distribution and performance analysis, as illustrated in Figure \ref{fig:ralph_cnot_simul_result}.
\begin{lstlisting}[language=python,style=mypython,caption= Analysis of the input-output map of the $\CNOT$ from Code \ref{code:cnot},captionpos=b,label=code:analyzer]
def post_process(output_state):
    '''postselected states are those containing one photon in the modes 
    {1,2} and the other in the modes {3,4}'''
    return (output_state[1] + output_state[2]) == 1 \
        and (output_state[3] + output_state[4]) == 1

cnot_processor = pcvl.Processor("SLOS", cnot)
cnot_processor.set_postprocess(post_process)

states = {
    pcvl.BasicState([0, 1, 0, 1, 0, 0]): "00",
    pcvl.BasicState([0, 1, 0, 0, 1, 0]): "01",
    pcvl.BasicState([0, 0, 1, 1, 0, 0]): "10",
    pcvl.BasicState([0, 0, 1, 0, 1, 0]): "11"
}

analyzer = pcvl.algorithm.Analyzer(cnot_processor, states)
analyzer.compute(expected={"00": "00", "01": "01", "10": "11", "11": "10"})
pcvl.pdisplay(analyzer, output_format=pcvl.Format.LATEX)
print("=> performance=%s, fidelity=%.1f%%" %
      (pcvl.simple_float(analyzer.performance)[1], analyzer.fidelity*100))
\end{lstlisting}

\begin{figure}[ht]
    \centering
    \begin{tabular}{rrrrr}
\hline
    &   00 &   01 &   10 &   11 \\
\hline
 00 &    1 &    0 &    0 &    0 \\
 01 &    0 &    1 &    0 &    0 \\
 10 &    0 &    0 &    0 &    1 \\
 11 &    0 &    0 &    1 &    0 \\
\hline
\end{tabular}

    \texttt{=> performance=1/9, fidelity=100.0\%}\hfill
    \caption{Output of Code \ref{code:analyzer}}
    \label{fig:ralph_cnot_simul_result}
\end{figure}

%% file: s4_research.tex
\section{\perceval in Action}
\label{sec:action}

In this section we provide examples of the \perceval software in use. We reproduce several photonic experiments that implement important quantum
algorithms and then demonstrate a photonic quantum machine learning algorithm on a simulated photonic quantum processor.

\subsection{The Hong-Ou-Mandel Effect}
\label{sec:hom}

\subsubsection{Introduction}

The Hong-Ou-Mandel (HOM) effect \cite{hong_measurement_1987} is an interference effect between pairs of indistinguishable photons, which, when incident on a balanced beam splitter via the respective input modes, will bunch and emerge together in either of the output modes with equal probability.
In 2002, Santori et al.\ demonstrated the HOM effect with a unique source of single photons \cite{santori2002indistinguishable},
evidencing the indistinguishability of consecutive photons and paving the way to many
developments in experimental quantum optics.

\begin{figure}[ht]
	\centering
	\includegraphics[width=\textwidth/2]{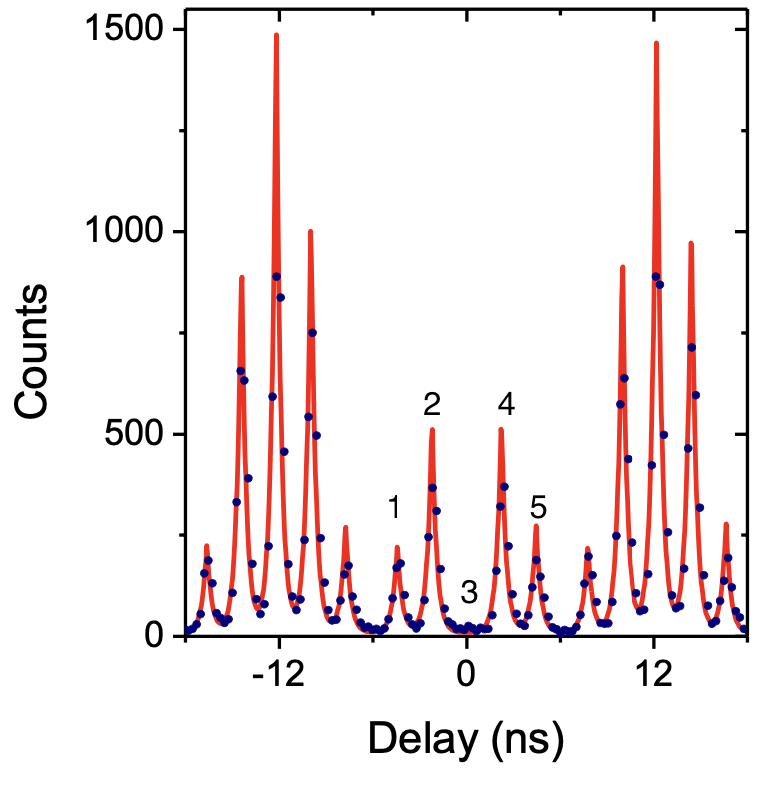}
    \caption{Measured correlation histogram between the HOM outputs in linear scale \cite{giesz2015cavity}. The red line shows the theoretical histogram for a perfect two-photon interference.}
	\label{fig:giesz2015cavity}
\end{figure}

In this experiment, ``{\it two pulses separated by $2$ ns and containing $0$ or
$1$ photons, arrive through a single-mode fibre. The pulses are interfered with
each other using a Michelson-type interferometer with a ($2$ ns$+\Delta{t}$)
path-length difference. [\dots] The interferometer outputs are collected by photon counters, and the resulting electronic signals are correlated using a time-to-amplitude converter followed by a multi-channel analyser card, which generates a histogram of the relative delay time $\tau=t_2-t_1$ between a photon detection at one counter ($t_1$) and the other ($t_2$)}'' \cite{santori2002indistinguishable}. An equivalent experiment was carried out in \cite{giesz2015cavity} with similar results, presented in Figure \ref{fig:giesz2015cavity}.

\subsubsection{\perceval Implementation}

The implementation on \perceval uses a simple circuit composed of a beam splitter
(corresponding to the main beam splitter of \cite[Figure 3(a)]{santori2002indistinguishable}),
the lower output mode of which then passes through a $1$-period
time-delay,
and then a second beam splitter
on the modes where the interference happens. 

\begin{figure}[ht]
	\centering
	\begin{subfigure}[t]{0.45\textwidth}
		\centering
		\includegraphics[width=\textwidth]{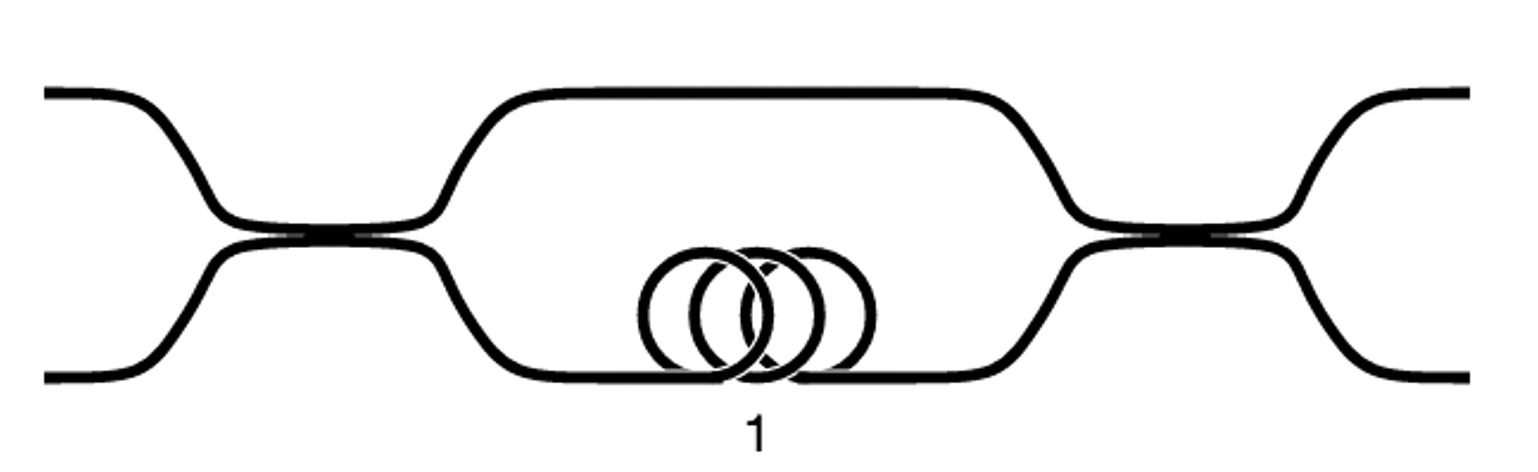}
		\caption{}
	\end{subfigure}
	\quad
	\begin{subfigure}[t]{0.45\textwidth}
		\centering
		\includegraphics[width=\textwidth]{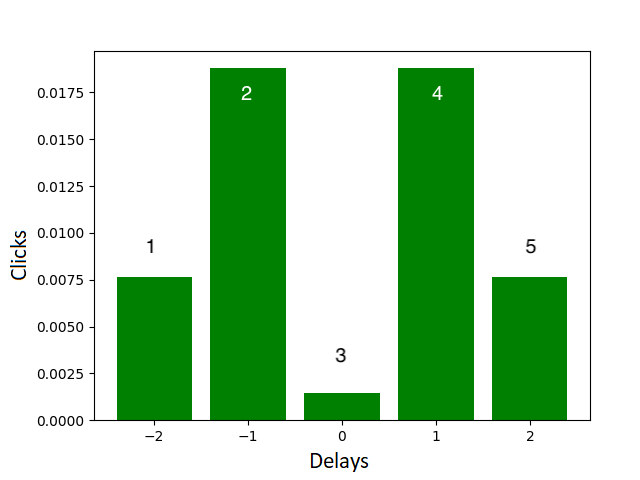}
		\caption{}
	\end{subfigure}
  \caption{Reproduction of the experiments of
  \cite{santori2002indistinguishable} and \cite{giesz2015cavity} in \perceval.
  \emph{(a)} The circuit as constructed in \perceval. \emph{(b)} Time
  simulation is run with \perceval and the difference between two consecutive
  detections on both arms is plotted. An imperfect source emitting a photon with a 30\% probability, and with probability
  $1\%$ of generating two photons at each cycle has been simulated. This slight $g_2^{(0)}$ imperfection explains the nonzero value at $\tau=0$.}
	\label{fig:perceval-timedelay}
\end{figure}

The circuit and result are presented in Figure \ref{fig:perceval-timedelay}.
The resolution is less fine-grained than the original experiments due to the
absence of modelling of photon length in the current implementation of
\perceval. However, we clearly recognise the same distinctive peaks in the
relative amplitude. The very small coincidence rate at $\tau=0$ is the
signature of photon interference.
The implementation of the algorithm is provided in Appendix.

\subsection{Boson Sampling}
\label{sec:bs}

\subsubsection{Introduction}
Boson Sampling\footnote{The technical introduction to the problem here draws heavily on a previous work by a subset of the authors of this paper \cite{MM22}.} is a \emph{sampling} problem originally proposed by Aaronson and Arkhipov \cite{aaronson_computational_2011}.
We give a detailed description of the problem below.
It essentially consists of sampling from the probability distribution of outcome detection coincidences when single photons are introduced into a random linear optical circuit.

Let $m, n \in \mathbb{N}^{*}$ be positive integers with $m \geq n$. Let
$\mathcal{S}_{m,n}$ be the set of all possible tuples $(s_1,\dots,s_m)$ of $m$
non-negative integers $s_i \in \mathbb{N}$, with $\sum_{i=1}^m s_i=n$. Let $U$
be an $m \times m$ Haar-random unitary matrix.
From $U$ we construct an $n \times n$ matrix $U_{T,S}$ as follows.
For a given $S=(s_1,\dots,s_m) \in \mathcal{S}_{m,n}$,
first construct an $n \times m$ matrix $U_{S}$ by copying $s_i$ times the $i$\textsuperscript{th} row of $U$.
Then, for a given (fixed) $T=(t_1,\dots,t_m) \in \mathcal{S}_{m,n}$, construct $U_{T,S}$ from $U_S$ by taking $t_j$ copies of the $j$\textsuperscript{th} column of $U_S$. Let $\mathsf{Perm}(U_{T,S})$ be the \emph{permanent} \cite[Chapter~7]{brualdi1991combinatorial} 
of $U_{T,S}$, and let
\begin{equation}
 \label{eqperm}
 P(S)=\frac{|\mathsf{Perm}(U_{T,S})|^2}{s_1 !\dots s_m!t_1! \dots t_m!}.
\end{equation}
It can be shown that $P(S) \in [0,1]$, and that the set $D_{U}=\{P(S) \mid S \in \mathcal{S}_{m,n}\}$ is a probability distribution over \emph{outcomes} $S$ \cite{aaronson_computational_2011}.

Let $\varepsilon \in [0,1]$ be a given \emph{precision}. Approximate Boson Sampling can then be defined as the problem of sampling outcomes $S$ from a probability distribution $\tilde{D}$ such that
\begin{equation}
 \label{bosonsampling}
 \|\tilde{D}-D_{U}\| \leq \varepsilon,
\end{equation}
where $\|.\|$ denotes the total variational distance between two probability
distributions. Exact Boson Sampling corresponds to the case in which $\varepsilon=0$.

For $m \gg n^2$, it was shown in \cite{aaronson_computational_2011} that:
\begin{itemize}
    \item No classical algorithm can solve Exact Boson Sampling in time $\mathsf{poly}(n)$,
    \item No classical algorithm can solve Approximate Boson Sampling in time $\mathsf{poly}(n,\frac{1}{\varepsilon})$, unless some widely accepted complexity theoretic conjectures turn out to be false.
\end{itemize}
These hardness results are closely related to the hardness of computing the permanent of matrices \cite{valiant1979complexity,aaronson_computational_2011}. 

On the other hand, Boson Sampling can be solved efficiently in the exact case on a photonic quantum device which is noiseless \cite{aaronson_computational_2011}, while the approximate version only requires sufficiently low noise levels \cite{AB16,Arkhipov15,KK14}.
This is done by passing $n$ identical single photons through a lossless
$m$-mode {universal} linear optical circuit \cite{Reck94,Clements16} configured in such a way that it implements the desired Haar-random unitary transformation $U$.  
Universal here is meant in the sense of photonic unitaries and refers to the
ability of the circuit to generate any arbitrary unitary evolution on the creation
operators of the modes. A universal linear optical circuit can be configured
to implement {any} $m \times m$ unitary chosen from the Haar measure,
for example using the recipe of \cite{Haar17}.
The input configuration of single photons corresponds to a tuple $T$.
We then measure the output modes of the circuit using perfect single photon detectors,
and will obtain the output configuration tuple $S$ according to the distribution $D_U$ \cite{aaronson_computational_2011}. The detectors should be photon-number resolving in general, but not necessarily when working in the no-collision regime ($n^2<<m$) \cite{aaronson_computational_2011}.

\subsubsection{\perceval Implementation\texorpdfstring{\protect\footnotemark}{}}

\footnotetext{This implementation is accompanied by a \href{https://perceval.quandela.net/docs/notebooks/Boson\%20Sampling.html}{notebook}.}

Boson Sampling with single photons in a $60$ mode linear optical circuit with up to $14$-photon coincidences at the outputs was reported in \cite{wang_boson_2019}.
In this Section, we report the results of a noiseless Boson Sampling simulation performed with \perceval for $n=14$  single photons and $m=60$ modes. For this Boson Sampling, the total number of possible output states is  $M=\binom{m+n-1}{n}=\binom{73}{14} \approx 10^{14}$ \cite{aaronson_computational_2011}.

For an input state of $14$ photons in $14$ arbitrarily chosen modes $k_1,...,k_{14} \in \{0,...,59\}$, we generated $300$ Haar random $60 \times 60$  unitaries, and performed for each of these unitaries $5 \times 10^6$ runs of Boson Sampling, where each run consists of sampling a single output. In total, we collected $1.5 \times 10^9$ samples.

The classical algorithm for Boson Sampling integrated into \perceval and which was used for this simulation is that of Clifford and Clifford \cite{clifford_classical_2018}. Note that faster versions of this algorithm have been found by the same authors in the case where $m$ is proportional to $n$ \cite{clifford2020faster}. Integration of the algorithm of \cite{clifford2020faster} into \perceval is a subject of on-going work.
The simulations were performed on a 32-core 3.1GHz Intel Haswell processor, at the rate of 8547 runs (samples) per second. It took roughly $2$ days to collect $1.5 \times 10^9$ samples.

Brute-force certification that our simulation has correctly implemented Boson Sampling would require, for each of the 300 Haar random unitaries, calculation of approximately $10^{14}$ permanents (to get the ideal probability distribution of the Boson Sampler), then performing $\approx 10^{14}$ runs of Boson Sampling using our simulation, in order to get the distribution of the simulated Boson Sampler to within the desired accuracy.
Clearly, this task is computationally intractable. In order to get around this issue,
while still having some level of confidence that our Boson Sampling simulation has indeed been implemented correctly, we have appealed to computing {partial} certificates \cite{MM22,shchesnovich2016universality}. These certificates are usually efficiently computable, but nevertheless can be used to rule out \emph{some} common adversarial strategies designed to \emph{spoof} Boson Sampling \cite{tichy2014stringent,walschaers2016statistical},
although they cannot provably rule out {every} possible adversarial spoofing strategy \cite{tichy2014stringent,shchesnovich2016universality,walschaers2016statistical,MM22}. 

The partial certificate we used here is the probability 
$P(K)$ that all $n$ input photons are measured in the first $K$ output modes of the Boson Sampler \cite{shchesnovich2016universality}, where $K \leq m$. We believe this choice of partial certificate is natural for benchmarking our simulation mainly for the following reasons. First, it can be computed straightforwardly from the output probabilities of Boson Sampling. Second, it relies on computing high order marginals, which are most likely difficult to compute efficiently classically \cite{aaronson_computational_2011}. Finally, for some values of $n,m$, and $K$, $P(K)$ can be computed to very good accuracy by using only a polynomial number of samples from the device \cite{shchesnovich2016universality}.

\begin{equation}
    \label{eqpk}
    P(K):=P(0_{K+1}, \dots, 0_{m})=\sum_{\{S \mid s_{K+1}=\dots=s_{m}=0\}}P(S),
\end{equation}
where, for a given $m \times m$ unitary $U$, $P(S)$ is computed as in Equation \ref{eqperm}. The average $\langle P(K) \rangle$  of $P(K)$ over the Haar measure
of $m \times m$ unitaries $U$
has been computed analytically in \cite{shchesnovich2016universality}:
\begin{equation}
    \label{eqhaarpk}
    \langle P(K) \rangle=\frac{K(K+1) \dots (K+n-1)}{m(m+1) \dots (m+n-1)}.
\end{equation}

We computed an estimate $\langle \tilde{P}(K) \rangle$ of $\langle P(K) \rangle$ by computing an estimate $\tilde{P}(K)$ of $P(K)$ using $5 \times 10^6$ samples for each of the 300 Haar random unitaries, then performing a uniform average over all these 300 unitaries. Our calculations for various values of $K$, together with the standard deviation of the distribution of $\tilde{P}(K)$  are presented  in Table~\ref{figtablebs}.

\renewcommand{\arraystretch}{1.2}

\begin{table}[ht]
\centering
\begin{tabular}{|c|c|c|c|c|c|}
\hline
$K$                            & $30$        & $40$       & $50$      & $55$      & $57$      \\ \hline
$\langle \tilde{P}(K) \rangle$ & $0.015\%$   & $0.54\%$   & $9.30\%$  & $32.03\%$ & $50.89\%$ \\ \hline
$\langle P(K) \rangle$         & $0.021\%$   & $0.65\%$   & $10.11\%$ & $33.33\%$ & $52.20\%$ \\ \hline
Std. Dev.                      & $0.00005\%$ & $0.0014\%$ & $0.015\%$ & $0.035\%$ & $0.045\%$ \\ \hline
\end{tabular}
\caption{Analytical values ($\langle P(K) \rangle$, and values computed with \perceval ($\langle \tilde{P}(K) \rangle$) of the probability that all $14$ photons gather in the first $K$ out of $60$ output modes, for various values of $K$.}
\label{figtablebs}
\end{table}

\renewcommand{\arraystretch}{1}

We observe a good agreement between the analytical values and the values computed with \perceval, in particular when the value of $K$ is close to 60.
This is to be expected, since these values of $K$ are closest to the regime in which $n(m-K) \ll m$,
where it has been shown \cite{shchesnovich2016universality} that $\langle P(K) \rangle$ is polynomially close to one in $n$ and $m$,
and therefore a $\mathsf{poly}(n,m)$ number of samples from the Boson Sampler is enough to compute $\tilde{P}(K)$ to good precision so as to allow a reliable certification.

As a final remark, note that \perceval can also simulate \emph{imperfect} Boson Sampling, where the imperfections integrated so far include photon loss and the possibility of multi-photon emissions. For further details we refer the reader to the  \href{https://perceval.quandela.net/docs/}{documentation}.

\subsection{Grover's Algorithm}
\label{subsec:grover}
\subsubsection{Introduction}
Grover's algorithm \cite{grover_fast_1996} is an $\bigO{\sqrt{N}}$-runtime quantum algorithm for searching an unstructured database of size $N$.
It is remarkable for providing a polynomial speedup over the best known classical algorithm for this task, which runs in time $\bigO{N}$. We now provide an intuitive explanation of how this algorithm works, similar to that which can be found in the \textit{Qiskit} \href{https://qiskit.org/textbook/ch-algorithms/grover.html}{documentation} or in \cite{roy2022deterministic}.

We work in the $n$-qubit Hilbert space $(\mathbb{C}^2)^{\otimes n}$, which allows us to treat a database of size $N=2^n$. In its standard formulation, the goal of Grover's algorithm is to look for an unknown computational basis state  $|T\rangle$, which we will call the {target} state, with $T \in \{0,1\}^n$.

Let $|R\rangle$ be the (normalised)  uniform superposition of all the $N-1$ computational basis states other than $|T\rangle$, which is orthogonal to $|T\rangle$. The uniform superposition $|S\rangle$ of all $N$ computational basis states can then be written as
\begin{equation}
|S\rangle:=\frac{1}{\sqrt{N}}|T\rangle+\frac{\sqrt{N-1}}{\sqrt{N}}|R\rangle. 
\end{equation}
Describing the problem setting in this way allows us to describe Grover's algorithm in a {geometric} picture. Indeed, we can now think of the states $|T\rangle$, $|R\rangle$, and $|S\rangle$ as lying in a 2-dimensional plane with a basis $\{|R\rangle,|T\rangle\}$. For example, $|S\rangle$ forms an angle $\theta=\mathsf{arcsin}(\frac{1}{\sqrt{N}})$ with respect to  $|R\rangle$ (or $\frac{\pi}{2}-\theta$ with respect to $|T\rangle$). 
 
Grover's algorithm relies on the application of two unitaries noted $U_O$ and $U_d$. The former is an \emph{oracle} unitary\footnote{An oracle is a unitary which the algorithm can apply without having knowledge about its internal implementation.} defined on computational basis states $|x\rangle$ as
 \begin{equation}
 U_O|x\rangle := 
 \begin{cases}
    -|x\rangle & \text{if}\quad x = T\\
    |x\rangle & \text{otherwise}
 \end{cases}
 \, ;
 \end{equation}
while the latter is the \emph{diffusion} unitary $U_d$ is defined as 
\begin{equation}
U_d:=2|S\rangle\langle S|-\mathbf{1}_n \, ,
\end{equation}
where $\mathbf{1}_n$ the $n$-qubit identity matrix.
On the uniform superposition, the unitary $U_O$ acts as 
 \begin{equation}
 U_O|S\rangle:=-\frac{1}{\sqrt{N}}|T\rangle+\frac{\sqrt{N-1}}{\sqrt{N}}|R\rangle \, .
 \end{equation}
Geometrically,
$U_O$ reflects the state $|S\rangle$ through the vector $|R\rangle$. Let's call the resulting reflected state $|S_0\rangle$.  Then,  $U_d$  performs a reflection of $|S_0\rangle$ through the vector $|S\rangle$. Let the resulting reflected state be denoted $|S_{O,d}\rangle$. One can see that these transformations keep the states in the same 2D-plane, and that $|S_{O,d}\rangle$ forms an angle of $2\theta$ with respect to $|S\rangle$ (or $\frac{\pi}{2}-3\theta$ with respect to $|T\rangle$). The state after applying $U_dU_O$ is thus  closer to $|T\rangle$ than the initial state $|S\rangle$ was. 
  
The reflections that we described correspond to one iteration of Grover's algorithm. The full algorithm first creates the state $|S\rangle$ by applying $n$ Hadamard gates $\Ha^{\otimes n}$ to an input state $|0\rangle^{\otimes n}$. It then transforms the state $|S\rangle$ into $|T\rangle$ by applying $k$ times the unitary $U_dU_O$ to $|S\rangle$. Following the geometric reasoning above, the required number of iterations $k$ is given by:
 \begin{equation}
  \label{eqgroverit}
  k \approx \ceil*{\frac{\frac{\pi}{2}-\theta}{2\theta}}=\ceil*{\frac{\pi}{4 \mathsf{arcsin}(\frac{1}{\sqrt{N}})}-\frac{1}{2}} \approx \bigO{\sqrt{N}},
 \end{equation}
where $\lceil \cdot \rceil$ is the ceiling function.

As a closing remark, let us mention that the original version of Grover's algorithm presented here is non-deterministic, albeit with a probability of success approaching one exponentially quickly with increasing $N$ \cite{grover_fast_1996}. Fully deterministic versions of Grover's algorithm can be obtained, for example by using different oracles and diffusion unitaries than those used here \cite{long2001grover}, or by using two different types of diffusion unitaries while keeping the same $U_O$ \cite{roy2022deterministic}.

\subsubsection{\perceval Implementation\texorpdfstring{\protect\footnotemark}{}}

\footnotetext{This implementation is accompanied by a \href{https://perceval.quandela.net/docs/notebooks/2-mode\%20Grover\%20algorithm.html}{notebook}.}

Here we reproduce the photonic demonstration of Grover's algorithm of \cite{kwiat_grovers_2000}. It uses the spatial and polarisation degrees of freedom to implement a mode realisation of the algorithm with two spatial modes. We spell out the correspondence between the marked database element and the associated quantum state in Table \ref{tab:marked}.

\begin{table}[htp]
\centering
\begin{tabular}{|c|c|c|c|c|c|}
\hline
Marked database element      & Quantum state      \\ \hline
$00$ & $\ket{0,P:H}$   \\ \hline
$01$        & $\ket{0,P:V}$    \\ \hline
$10$ & $\ket{P:H,0}$   \\ \hline
$11$        & $\ket{P:V,0}$    \\ \hline
\end{tabular}
\caption{Equivalence between database elements and states in \perceval.}
\label{tab:marked}
\end{table}

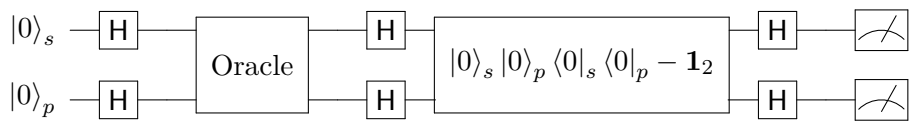
\begin{figure}[ht]
    \centering
    \[
    \Qcircuit @C=1em @R=1em {
        \lstick{\ket{0}_s} & \gate{\Ha} & \qw  &  \multigate{1}{\text{Oracle}} & \qw & \gate{\Ha} & \multigate{1}{\ket{0}_s\ket{0}_p\bra{0}_s\bra{0}_p-\mathbf{1}_2} & \gate{\Ha} & \qw & \meter{}  \\
        \lstick{\ket{0}_p} & \gate{\Ha} & \qw  &  \ghost{\text{Oracle}} & \qw  & \gate{\Ha} & \ghost{\ket{0}_s\ket{0}_p\bra{0}_s\bra{0}_p-\mathsf{1}_2} &  \gate{\Ha}  & \qw & \meter{}  \\
    }
    \]
    \caption{Quantum circuit implementing Grover's algorithm using the spatial ($s$) and polarisation ($p$) degrees of freedom. H denotes a Hadamard gate.
    The first set of Hadamard gates creates a uniform superposition over all states. Subsequently, the oracle is applied on the superposition, followed by a Grover diffusion operation coupled to detection devices.}
    \label{fig:quantum_circuit_grover}
\end{figure}

\paragraph{Quantum Circuit.}
Figure \ref{fig:quantum_circuit_grover} represents the quantum circuit that implements Grover's algorithm on two qubits, encoded over two spatial modes and their polarisation modes as described in Table \ref{tab:marked}. It features a state initialisation stage creating a uniform quantum superposition over all states, an oracle stage, and a diffusion stage.
Replacing $N=4$ in Equation \ref{eqgroverit} gives $k \approx \frac{\pi}{4.\frac{\pi}{6}}-\frac{1}{2}=1.$ Thus, one application of the oracle and diffusion gates suffices.

\begin{figure}[ht]
    \centering
    \includegraphics[width=15cm]{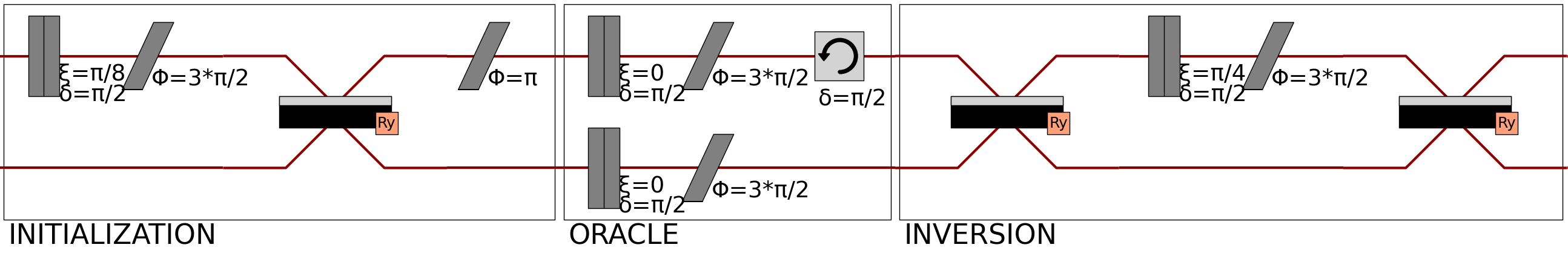}
    \caption{Linear optical circuit implementing Grover's algorithm. The oracle is set here to mark the element '00', which encodes the state $\ket{0,{P:H}}$. Because the circuit has been compiled, the initialisation, oracle and inversion stages do not strictly apply the expected operation. The half-wave plates in this figure are followed by a $-\pi/2$ phase shift to match the half-wave plate definition of Kwiat et al.\ The figure has been generated using \perceval.}
    \label{fig:circuit_grover}
\end{figure}

\paragraph{Linear Optical Circuit.}
The linear optical circuit we will use in our simulation is shown in Figure \ref{fig:circuit_grover}. This circuit was experimentally realised by Kwiat et al.\ \cite{kwiat_grovers_2000} using bulk (\ie free-space) optics and compiled to limit the number of optical elements introduced in the experimental setup, by moving and combining operations like phase shifts. 

\begin{figure}[ht]
    \centering
    \includegraphics[width=15cm]{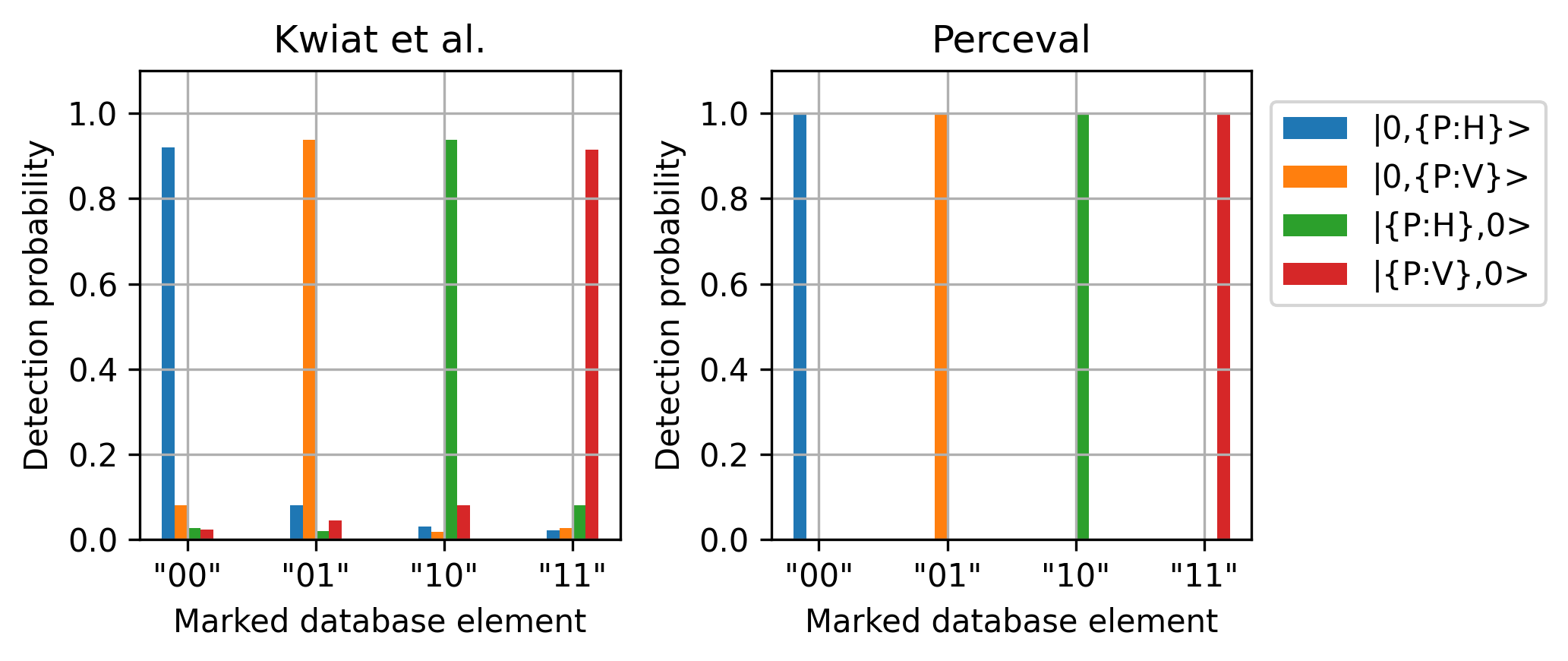}
    \caption{Left: experimental results from Kwiat et al.\ \cite{kwiat_grovers_2000}. Right: results of the simulation in \perceval. The simulated results are as expected and thus match the results obtained in the experiment (up to experimental error).}
    \label{fig:grover_results}
\end{figure}

\paragraph{Simulation.}
Here, we simulate in \perceval the linear optical circuit of \cite{kwiat_grovers_2000} (see Figure \ref{fig:circuit_grover}), a photonic realisation of the quantum circuit in Figure \ref{fig:quantum_circuit_grover}, which implements Grover's algorithm for marked elements $00$, $01$, $10$ and 11. The results are displayed in Figure \ref{fig:grover_results}, along with the experimental results of Kwiat et al.\ \cite{kwiat_grovers_2000}.

\subsection{Shor's Algorithm}
\label{sec:shor}
\subsubsection{Introduction}

The problem of factoring an integer is thought to be in $\NP$\emph{-intermediate} and the best known classical algorithms only achieve sub-exponential running times. Its classical complexity is well studied since it has been used as the basis of the most widely adopted encryption scheme, Rivest-Shamir-Adleman (RSA) \cite{rivest2019method}, where the secret key consists in two large primes $p$ and $q$, while their product $N = pq$ is the corresponding public key. In this context, Shor's algorithm \cite{shor_algorithms_1994} greatly boosted interest in quantum algorithms by showing that such composite numbers can in fact be factored in polynomial time on a quantum computer.

\paragraph{From Period-Finding to Factoring (Classically).} 
We start by assuming that there exists a period-finding algorithm for functions over integers given as a black-box. This algorithm is used to find the order of integers $a$ in the ring $\Z_N$, \ie the smallest value $r$ such that $a^r \equiv 1 \pmod N$. In particular, the values of $a$ are sampled at random until the associated $r$ is even.

Once such an $a$ has been found,\footnote{Values of $a$ that verify this property are common in $\Z_N$.} we have that $a^r-1=(a^{\frac{r}{2}}-1)(a^{\frac{r}{2}}+1)$ is a multiple of $N$ but not $a^{\frac{r}{2}}-1$ (otherwise the period would be $r/2$). We test whether $a^{\frac{r}{2}}+1$ is a multiple of $N$ and if so restart the procedure.\footnote{It can be proven that the probability of not restarting at this step is high, and on average the algorithm needs to be repeated only once.} Otherwise, we have found a multiple of one of the prime factors $p,q$ since $a^{\frac{r}{2}}+1$ divides a multiple of $N$. Taking the greatest common divisor (GCD) of $a^{\frac{r}{2}}+1$ and $N$ easily yields this prime factor. Indeed, the GCD can be found efficiently classically for example by using Euclid's algorithm.

In summary, all the steps described are basic arithmetic operations, simple to implement on a classical machine, and most of the complexity is hidden in the period-finding algorithm.

\paragraph{Shor's Quantum Period-Finding Algorithm.}
The key contribution of Shor's work was to show that the period-finding algorithm can be done in polynomial time on a quantum computer. 

Indeed, finding the order for a value $a$ can be reduced to a problem of phase estimation for the unitary $U_a$ implementing the Modular Exponentiation Function (MEF) $x \rightarrow a^x \pmod N$ on computational basis states. It can be shown that all eigenvalues of these operators are of the form $e^{\frac{2i\pi k}{r}}$ for an integer $k$ and the value of interest $r$. The phase estimation algorithm uses as basis the Quantum Fourier Transform and its inverse, along with controlled versions of unitaries of the type $U_{a^{2^p}}$ up to $2^p \approx N^2$.\footnote{Using this family of unitaries reduces the precision required of the phase estimation procedure from $1/N^2$ to constant.} This part of the circuit is specific to the value of $a$ and $N$ and can therefore be optimised once they have been chosen. Although the most costly part of the algorithm in terms of gates, the unitaries $C$-$U_{a^{2^p}}$ (controlled $U_{a^{2^p}}$ gates) can still be implemented in polynomial time on a quantum computer, making the overall procedure efficient as well.

\subsubsection{\perceval Implementation\texorpdfstring{\protect\footnotemark}{}}

\footnotetext{This implementation is accompanied by a \href{https://perceval.quandela.net/docs/notebooks/Shor\%20Implementation.html}{notebook}.}

Here we reproduce the photonic realisation of Shor's algorithm from \cite{politi_shors_2009}.

\paragraph{Quantum Circuit.}

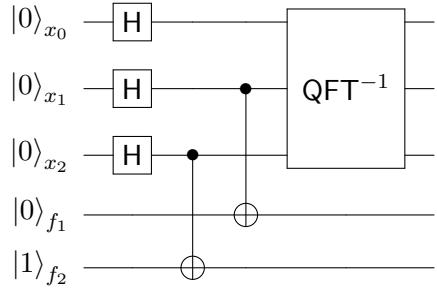
\begin{figure}[ht]
    \centering
    \[
    \Qcircuit @C=1em @R=1em {
        \lstick{\ket{0}_{x_0}} & \gate{\Ha} & \qw      & \qw      & \multigate{2}{\mathsf{QFT}^{-1}} & \qw \\
        \lstick{\ket{0}_{x_1}} & \gate{\Ha} & \qw      & \ctrl{2} & \ghost{\mathsf{QFT}^{-1}}        & \qw \\
        \lstick{\ket{0}_{x_2}} & \gate{\Ha} & \ctrl{2} & \qw      & \ghost{\mathsf{QFT}^{-1}}        & \qw \\
        \lstick{\ket{0}_{f_1}} & \qw        & \qw      & \targ    & \qw                              & \qw \\
        \lstick{\ket{1}_{f_2}} & \qw        & \targ    & \qw      & \qw                              & \qw
    }
    \]
    \caption{Quantum circuit implementing Shor's algorithm for $N = 15$ and $a = 2$.}
    \label{fig:quantum_circuit_shor}
\end{figure}

The quantum circuit shown in Figure \ref{fig:quantum_circuit_shor} for factoring $N = 15$ using parameter $a = 2$, whose order is $r=4$.
It acts on 5 qubits labelled $x_0, x_1, x_2$ (for the top three) and $f_1, f_2$ (for the bottom two). The $\CNOT$ gates apply a version of the MEF which has been optimised for this specific value of $a$ to the qubits $x_i$ in superposition, storing the outcome in qubits $f_i$. The outcome is given by measuring the qubits $x_1, x_2$, followed by classical post-processing.

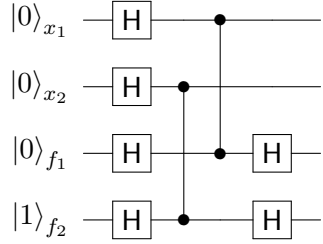
\begin{figure}[ht]
    \centering
\[
\Qcircuit @C=1em @R=1em {
    \lstick{\ket{0}_{x_1}} & \gate{\Ha} & \qw          & \ctrl{2}     & \qw        & \qw \\
    \lstick{\ket{0}_{x_2}} & \gate{\Ha} & \ctrl{2}     & \qw          & \qw        & \qw \\
    \lstick{\ket{0}_{f_1}} & \gate{\Ha} & \qw          & \control \qw & \gate{\Ha} & \qw \\
    \lstick{\ket{1}_{f_2}} & \gate{\Ha} & \control \qw & \qw          & \gate{\Ha} & \qw
}
\]
    \caption{Simplified Shor quantum circuit.}
    \label{fig:simpl-shor}
\end{figure}

\paragraph{Linear Optical Circuit.} 
Since qubit $x_0$ remains unentangled from the other qubits it can be removed from the optical implementation of this circuit. Furthermore, the $\CNOT$ gates on $x_i, f_i$ can be realised as
\begin{equation}
    \CNOT_{1,2} = \Ha_2 \circ \CZ_{1,2} \circ \Ha_2,
\end{equation}
where the indices denote the qubits to which the gate is applied (in the case of $\CNOT$, the first qubit is the control). Finally, the inverse Quantum Fourier Transform can be performed via classical post-processing, and so does not need to be implemented as a quantum gate in the circuit. The circuit after these simplifications is given in Figure \ref{fig:simpl-shor}.

The expected output state of the circuit above is
\begin{equation}
\label{eq:shor-out}
\frac{1}{2} \left ( |0\rangle_{x_1}|0\rangle_{f_1} + |1\rangle_{x_1}|1\rangle_{f_1} \right ) \otimes \left ( |0\rangle_{x_2}|1\rangle_{f_2} + |1\rangle_{x_2}|0\rangle_{f_2} \right ).
\end{equation}
We work with path encoded qubits, as in \cite{politi_shors_2009}. With path encoding, each $\Ha$ gate in the quantum circuit is implemented with a beam splitter with reflectivity $R=1/2$ between the two paths corresponding to the qubit. In our implementation in \perceval, phase shifters are added to properly tune the phase between each path.

$\CZ$ gates are implemented with $3$ beam splitters with reflectivity $R=2/3$ acting on $6$ modes: one inner beam splitter creates interference between the two qubits, and two outer beam splitter balance detection probability using auxiliary modes.
This optical implementation succesfully yields the output state produced by a $\CZ$ gate with probability $1/9$; otherwise it creates a dummy state, which can be removed by post-selection.

\begin{figure}[h]
    \centering
    \includegraphics[height=12cm]{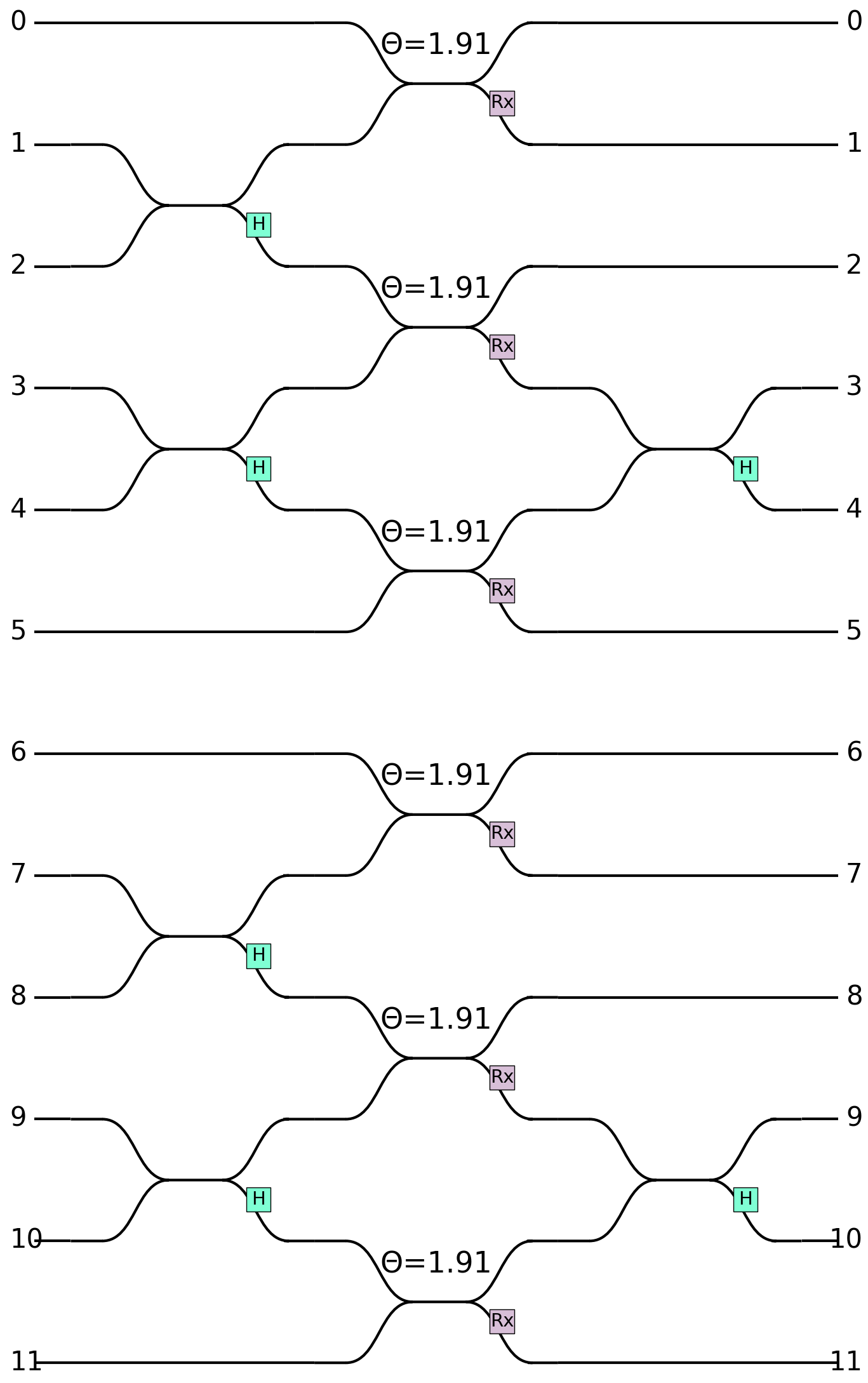}
    \caption{Linear optical circuit implementing Shor's algorithm. The $12$ modes are ordered from $0$ to $11$, from top to bottom. Modes $(1, 2)$, $(3, 4)$, $(7, 8)$, $(9, 10)$ encode qubits $x_1$, $x_2$, $f_1$, $f_2$ respectively. Modes $0, 5, 6, 11$ are the auxiliary modes for $\CZ$ gates.}
    \label{fig:circuit_shor}
\end{figure}

The circuit implemented in \perceval is illustrated in Figure \ref{fig:circuit_shor}.

\begin{figure}[h]
    \centering
    \includegraphics[height=6cm]{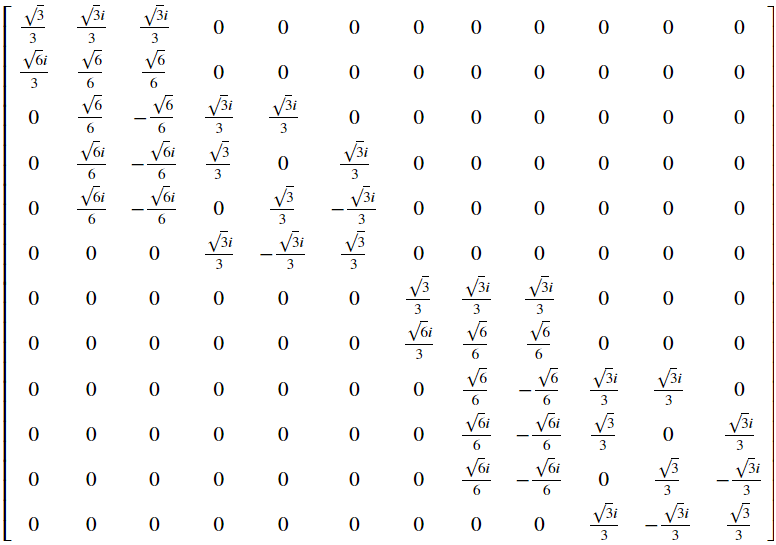}
    \caption{Unitary matrix associated with the optical circuit from Figure \ref{fig:circuit_shor}.}
    \label{fig:matrix_shor}
\end{figure}

The matrix associated to the optical circuit, giving the probability amplitude of each combination of input and output modes, is given in Figure \ref{fig:matrix_shor}.

\paragraph{Simulation.}
After entering the initial Fock state associated to the input qubit state $|0\rangle_{x_1}|0\rangle_{x_2}|0\rangle_{f_1}|1\rangle_{f_2},$ 
we plot the probability amplitudes of the output state, post-selected on Fock states corresponding to a qubit state with path encoding. The results are given in Figure \ref{fig:shor_amplitudes}.

\begin{figure}[htp]
    
    {\tt Output state amplitude: (post-selected on qubit states, not renormalized)\\
    |${\tt x}_{\tt 1}{\texttt{,x}}_{\tt 2}{\texttt{,f}}_{\tt 1}{\texttt{,f}}_{\tt 2}$>\\
    |0,0,0,0> 0j \\
    |0,0,0,1> (0.055555555555555566+0j)
\\
|0,0,1,0> 0j \\
|0,0,1,1> 0j \\
|0,1,0,0> (-0.055555555555555557-1.734723475976807e-18j)\\ 
|0,1,0,1> (-1.3877787807814457e-17-1.734723475976807e-18j)\\ 
|0,1,1,0> (-1.734723475976807e-18j+0j) \\ 
|0,1,1,1> 0j \\ 
|1,0,0,0> 0j \\ 
|1,0,0,1> (-1.3877787807814457e-17+0j)\\ 
|1,0,1,0> (-1.734723475976807e-18j+0j)\\ 
|1,0,1,1> (-0.055555555555555557-1.734723475976807e-18j) \\ 
|1,1,0,0> (1.3877787807814457e-17+0j) \\ 
|1,1,0,1> 0j\\
|1,1,1,0> (0.05555555555555558+0j)\\ 
|1,1,1,1> (1.3877787807814457e-17+5.204170427930421e-18j)}
    \caption{The output state amplitudes computed with \perceval, ${\tt j}^2=-1$.}
    \label{fig:shor_amplitudes}
\end{figure}

After re-normalisation we find that the output amplitudes computed with \perceval match the expected output state described in Equation \ref{eq:shor-out} up to numerical precision.

When decomposing the expected output state in the qubit basis, the qubit states with non-zero amplitude are
$\ket{0001}$, 
$\ket{0100}$, 
$\ket{1011}$ and 
$\ket{1110}$ for qubits $x_1,x_2,f_1,f_2$.
We plot the output distribution of the circuit, post-selected on these states, without re-normalisation; the result is presented in Table \ref{fig:shor_distribution}.

\renewcommand{\arraystretch}{1.2}

\begin{table}[htp]
\centering
\begin{tabular}{|c|c|c|c|c|}
\hline
                & $\ket{0,0,0,1}$ & $\ket{0,1,0,0}$ & $\ket{1,0,1,1}$ & $\ket{1,1,1,0}$\\ \hline
$\ket{0,0,0,1}$ & $0.003086$      & $0.003086$      & $0.003086$      & $0.003086$     \\ \hline
\end{tabular}
\caption{The output distribution for qubits $x_1,x_2,f_1,f_2$.}
\label{fig:shor_distribution}
\end{table}

\renewcommand{\arraystretch}{1}

The distribution obtained with \perceval is uniform over each outcome, which matches the expected distribution in \cite{politi_shors_2009}.

\paragraph{Outcome Distribution Interpretation.}

For each outcome, the values of qubits $x_2, x_1,$ $x_0$ (with $x_0 = 0$) represent a binary number between 0 and 7, here corresponding to $0, 4, 2, 6$ in the order of Table \ref{fig:shor_distribution}.
After sampling the circuit, obtaining outcomes $2$ or $6$ allows to successfully compute the order $r = 4$ \cite{politi_shors_2009}.
Obtaining outcome $0$ is an expected failure of the quantum circuit, inherent to Shor's algorithm.
Outcome $4$ is an expected failure as well, as it only allows to compute the trivial factors $1$ and $15$.

Since the distribution from Figure \ref{fig:shor_distribution} is uniform the circuit successfully yields a successful outcome with probability $1/2$. This probability can be amplified exponentially close to $1$ by sampling the circuit multiple times \cite{politi_shors_2009}.

\subsection{Variational Quantum Eigensolver}
\label{sec:vqe}
\subsubsection{Introduction}

The Variational Quantum Eigensolver (VQE) introduced by \cite{peruzzo_variational_2014} is an algorithm for finding eigenvalues of an operator. 
Applications range from finding ground state energies and properties of atoms and molecules to various combinatorial optimisation problems.

Since the eigenvector $|\psi^*\rangle$ associated to the smallest eigenvalue of $\mathcal{H}$ minimises the Rayleigh-Ritz quotient
\begin{equation}
\label{eq:RR-quotient}
\frac{\langle \psi^* | \mathcal{H} | \psi^* \rangle}{\langle \psi^* |\psi^* \rangle}, 
\end{equation}
the eigenvalue problem can be rephrased as a variational problem on this quantity.
The VQE algorithm uses as a sub-routine the Quantum Expectation Estimation (QEE) algorithm, developed in the same paper, whose task is precisely to compute the expectation value $\langle\mathcal{H}\rangle:=\langle  \psi| \mathcal{H}| \psi \rangle$ of Hamiltonian $\mathcal{H}$ for an input state $|\psi\rangle$ (here assumed to be normalised). 
Given an ansatz represented by tunable experimental parameters $\{\theta_i\}$, set to some initial values $\{\theta_i^\textrm{init}\}$, the VQE algorithm works by iterating a loop which first applies the QEE algorithm on a state generated using the initial parameters $\{\theta_i^\textrm{init}\}$. Then, based on the outcome, it tunes these parameters using a classical minimisation algorithm. This process is repeated until a termination condition specified by the classical minimisation algorithm is satisfied.

Let $n$ be the number of qubits of our system,  $\sigma_x$, $\sigma_y$, and $\sigma_z$ the single qubit Pauli $\sX$, $\sY$, and $\sZ$ matrices, and $1$ the single qubit identity matrix.

Any $n$-qubit Hamiltonian $\mathcal{H}$ can be decomposed as 
\begin{equation}
\mathcal{H}= \sum_{\sigma}h_{\sigma}\sigma,
\label{eqH}
\end{equation}
where $\sigma=\otimes^{n}_{i=1}\sigma_i$, $\sigma_{i} \in \{\sigma_x,\sigma_y,\sigma_z,1\}$, and  $h_{\sigma} \in \R$.

The expectation value $\langle \psi |\sigma| \psi \rangle$ of an $n$-qubit Pauli operator $\sigma$ with respect to a state $|\psi\rangle$ can be estimated efficiently by performing local measurements on $|\psi\rangle$ \cite{peruzzo_variational_2014}. Thus, if the number of terms in the sum in Equation \ref{eqH} is $\mathsf{poly}(n)$, then computing $\langle \mathcal{H} \rangle$ can be performed efficiently on a quantum device, as it reduces to computing $\mathsf{poly}(n)$  expectations of the form  $\langle \psi |\sigma| \psi \rangle$, each of which can be done efficiently. Unfortunately, various issues regarding the scalability of the VQE approach manifest with  increasing the system size. Notably, VQE circuits having a good expressivity \cite{du2020expressive} generally have a number of parameters scaling rapidly with the system size. Furthermore, in some cases,  the expectation value $\langle \psi |\sigma| \psi \rangle$ might be exponentially small in $n$, requiring, from standard statistical arguments \cite{hoeffding}, an exponential, and thus prohibitive, number of experimental samples to compute accurately.


\subsubsection{\perceval Implementation\texorpdfstring{\protect\footnotemark}{}}

\footnotetext{This implementation is accompanied by a \href{https://perceval.quandela.net/docs/notebooks/Variational\%20Quantum\%20Eigensolver.html}{notebook}.}

Here we use \perceval to reproduce the original photonic implementation of the VQE from \cite{peruzzo_variational_2014}. For small enough instances of the problem, \perceval is able to compute explicitly the complete state vector $|\psi\rangle$.  This allows us to skip the QEE subroutine and directly compute the mean value $\langle\mathcal{H}\rangle$.

\begin{figure}[ht]
    \centering
    \includegraphics[width=0.95\linewidth]{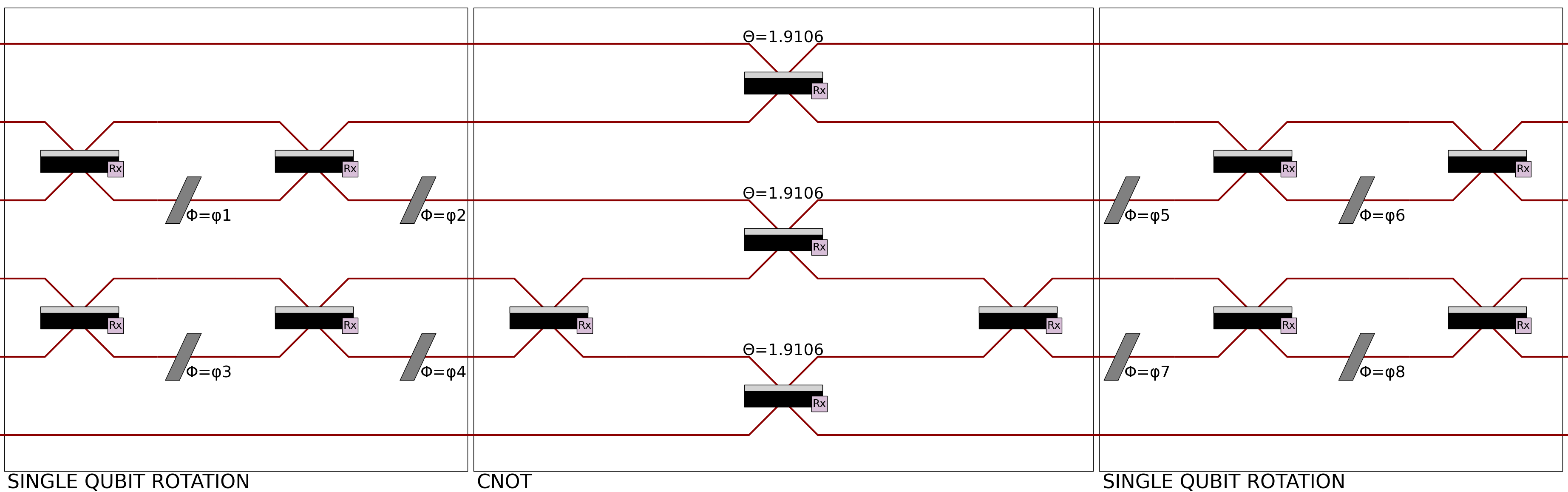}
    \caption{Linear optical circuit implementing the VQE algorithm. The $6$ modes are ordered from $0$ to $5$. Modes $(1,2)$ and $(3,4)$ encode the $2$ qubits. Modes $0$ and $5$ are the auxiliary modes for the $\CNOT$.}
    \label{fig:circuit_vqe}
\end{figure}

\paragraph{Linear Optical Circuit.}
We use the linear optical circuit of the original paper \cite{peruzzo_variational_2014} which was first introduced in \cite{noauthor_generating_2012}. The circuit has $6$ optical modes, consists of $13$ beam splitters, $8$ tunable phase shifters and $4$ single-photon detectors, and is shown in Figure \ref{fig:circuit_vqe}. This circuit is essentially a 2-qubit circuit where qubits $1$ and $2$ are path encoded respectively in mode pairs $(1,2)$
and $(3,4)$ (here mode numbering is from $0$ to $5$, from top to bottom). Modes $0$ and $5$ are auxiliary modes. The circuit consists of two parts. The first part being the $3$ central beam splitters in Figure \ref{fig:circuit_vqe}, together with the $2$ beam splitters to the left and right of these central beam splitters, and which act on modes $3$ and $4$. The $3$ central beam splitters have reflectivity $1/3$, whereas the two to the left and right of the central ones have reflectivity $1/2$.  These $5$ beam splitters are used, together with modes $0$ and $5$, to apply a $\CNOT$ gate to qubits $1$ and $2$ \cite{ralph_linear_2002}. This $\CNOT$ is successful with probability $1/9$ under post-selection. The second part of the circuit consists of the remaining $8$ beam splitters and phase shifters, and is used to implement arbitrary single qubit rotations on qubits $1$ and $2$. All $8$ phase shifters  have  tunable phase shifts, and all $8$ beam splitters have reflectivity $1/2$.

\paragraph{Simulation.}

Our strategy for the simulation of the VQE is as follows. We first compute the output state vector $|\psi\rangle$ of the linear optical circuit, which depends on the phase shifters parameters $(\phi_i)_{i \in {\{1, \dots, 8\}}}$ \cite{peruzzo_variational_2014}, for some random initial configuration of these parameters. This enables the evaluation of the Rayleigh-Ritz quotient in Equation \ref{eq:RR-quotient}. Then, we proceed to minimise this quantity by using the Nelder-Mead minimisation algorithm \cite{nelder1965simplex}.

\begin{figure}[ht]

	\centering
	\begin{subfigure}[t]{0.31\textwidth}
		\centering
		\includegraphics[width=\textwidth]{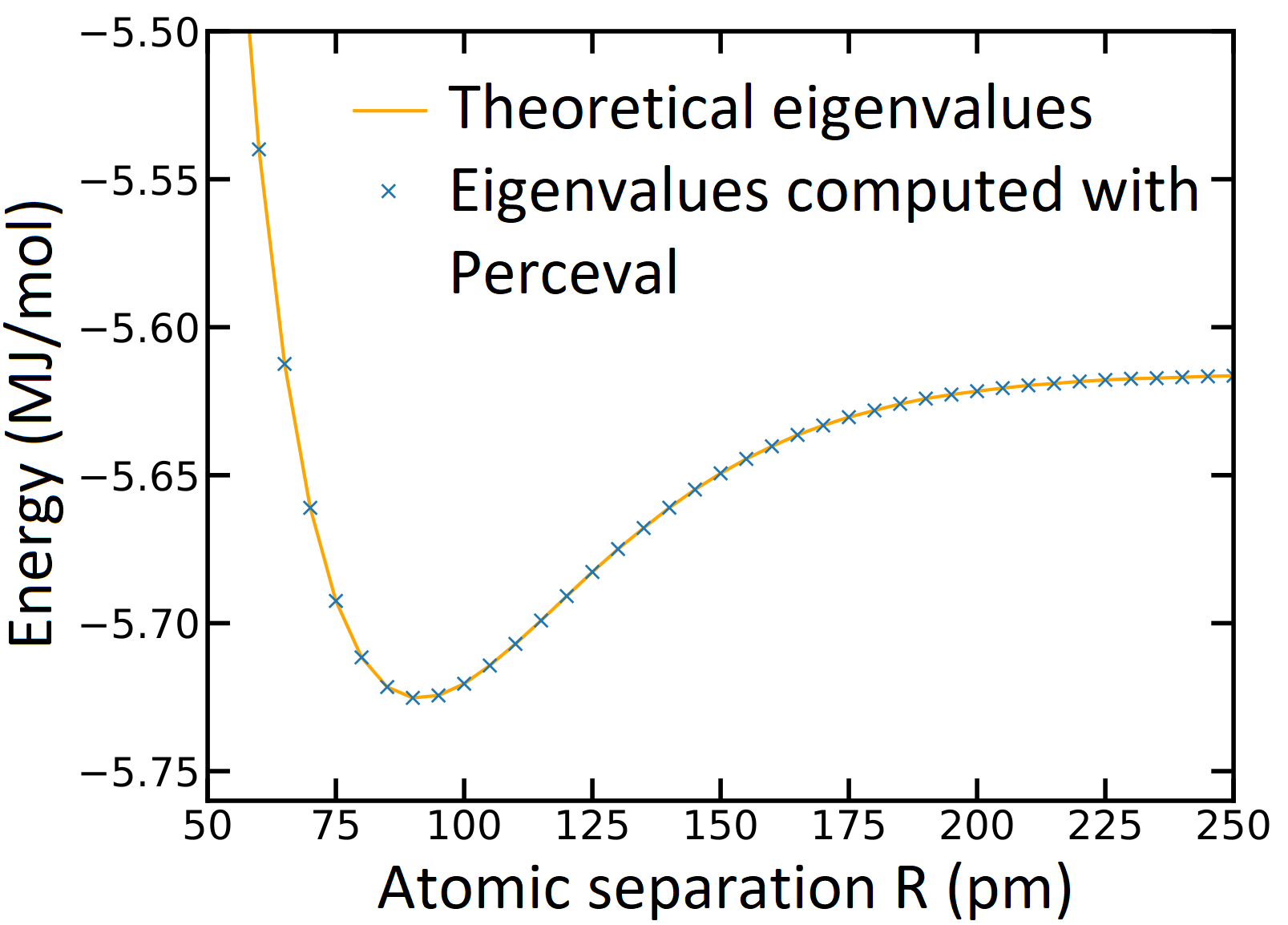}
		\caption{\perceval simulation of \\the Hamiltonian given in \cite{peruzzo_variational_2014}.}
		\label{fig:vqe1}
	\end{subfigure}
	\quad
	\begin{subfigure}[t]{0.31\textwidth}
		\centering
		\includegraphics[width=\textwidth]{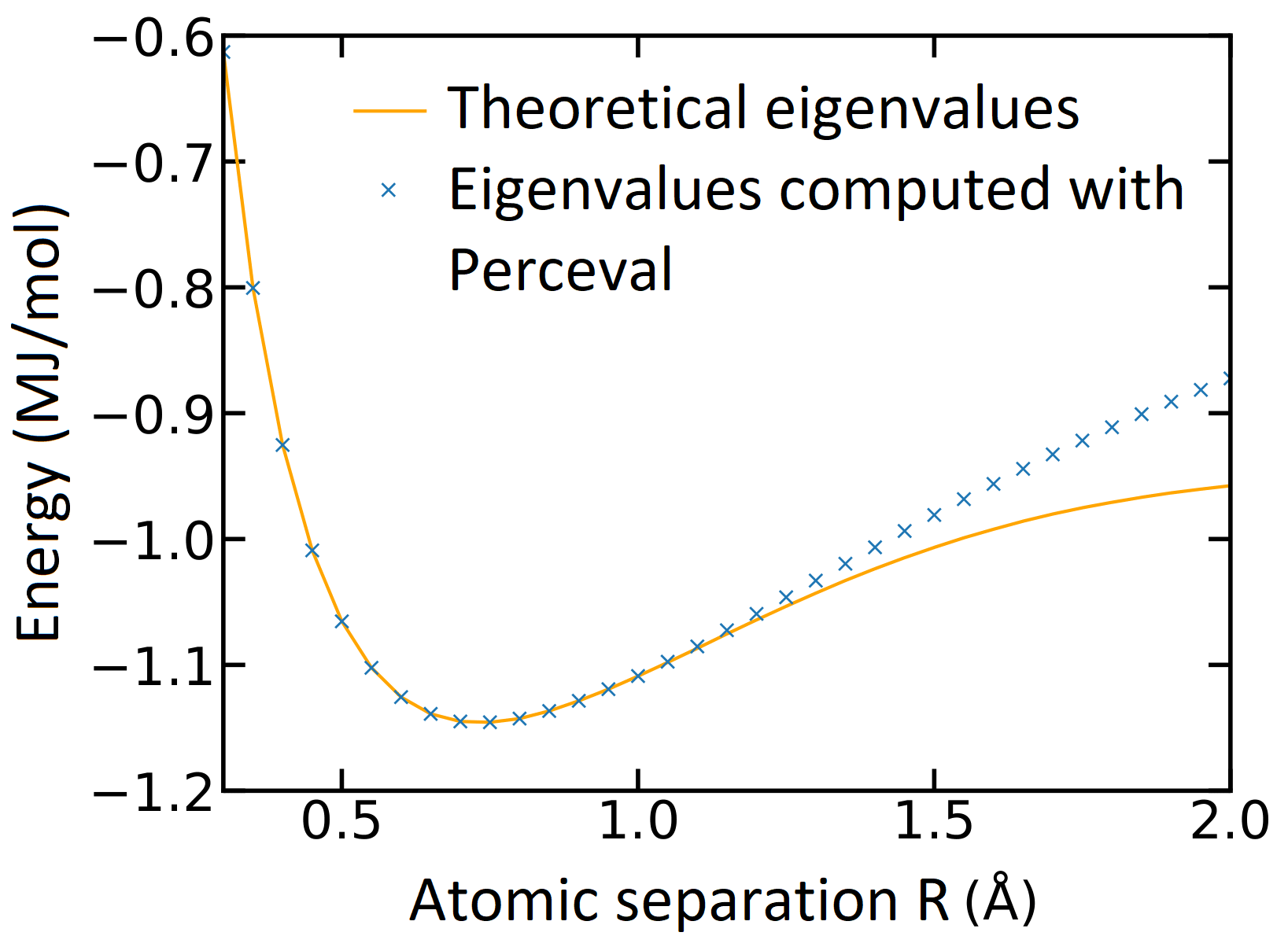}
		\caption{\perceval simulation of \\the Hamiltonian given in \cite{VQE2}.}
		\label{fig:vqe2}
	\end{subfigure}
	\quad
	\begin{subfigure}[t]{0.31\textwidth}
		\centering
		\includegraphics[width=\textwidth]{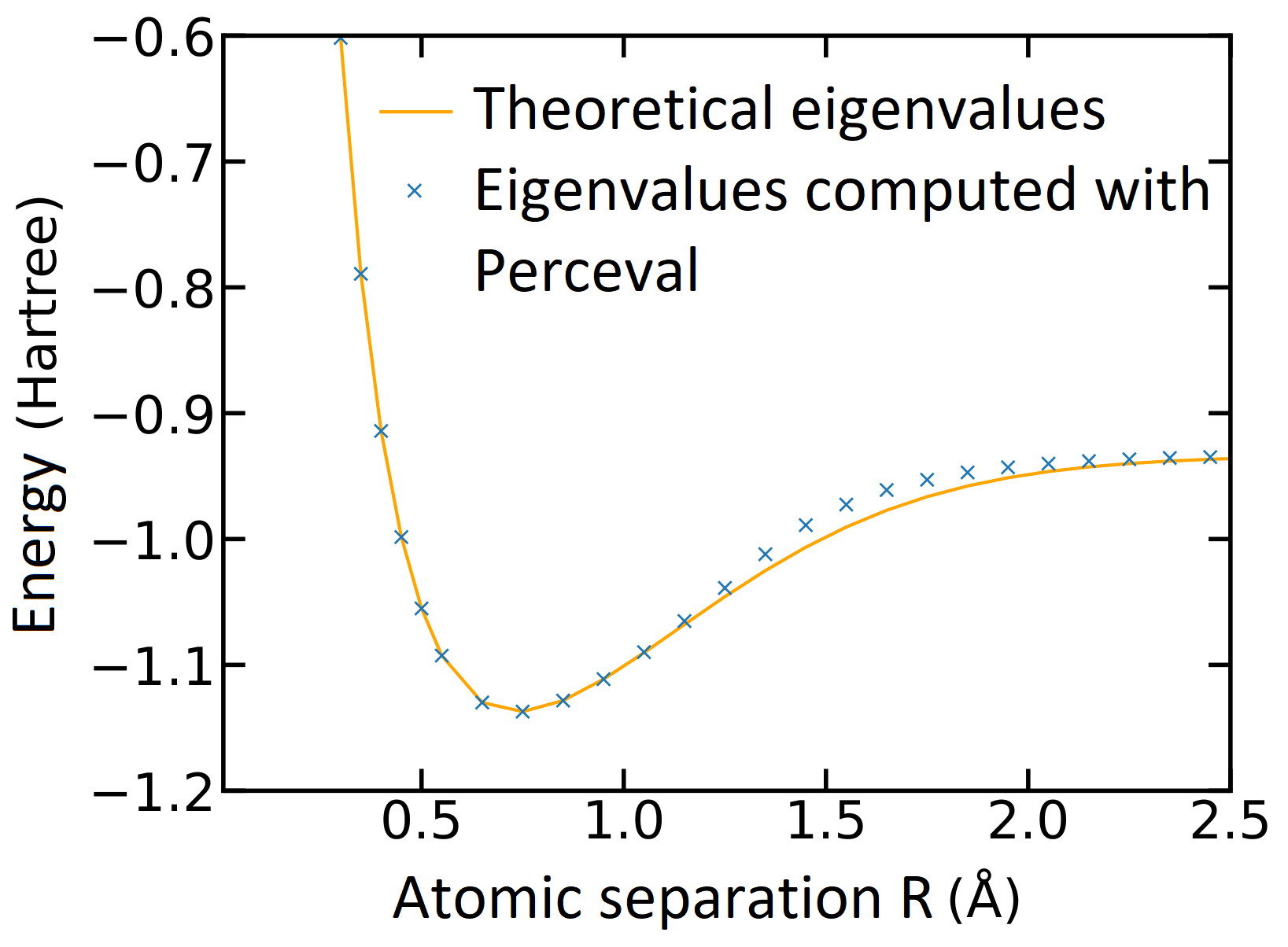}
		\caption{\perceval simulation of \\the Hamiltonian given in \cite{VQE3}.}
		\label{fig:vqe3}
	\end{subfigure}
  \caption{Simulations of ground-state molecular energies.}
	\label{figvqe}
\end{figure}

Using these techniques, we computed the ground-state molecular energies for the Hamiltonians of three different experiments \cite{peruzzo_variational_2014} \cite{VQE2} \cite{VQE3}. In all these experiments, we used the circuit of Figure \ref{fig:circuit_vqe} to compute the output states $|\psi\rangle$. In line with the VQE algorithm, at each iteration we produce a different $|\psi\rangle$ by tuning the angles of the phase shifters of the circuit in Figure \ref{fig:circuit_vqe}. To test the validity of our techniques, we computed  the theoretical ground-state molecular energies of the Hamiltonians of each of the 3 experiments. This was done with the \href{https://numpy.org/}{NumPy} \cite{harris2020array}  linear algebra package. A plot of these energies computed both theoretically and with \perceval is shown in 
Figure \ref{figvqe}.

The circuit of Figure \ref{fig:circuit_vqe} was originally used  in \cite{peruzzo_variational_2014} to compute the ground-state energies of the Hamiltonian in Figure \ref{fig:vqe1}. We observe a very good overlap between theoretical and computed energies, indicating that the \perceval simulation was succesful. In \cite{VQE2,VQE3}, different circuits than that of Figure \ref{fig:circuit_vqe} were used for computing the ground-state energies. Nevertheless, and in order to explore the expressivity \cite{du2020expressive} of the circuit of Figure \ref{fig:circuit_vqe}, we computed and plotted in \ref{fig:vqe2},\ref{fig:vqe3} the results of ground-state energy computations of the Hamitonians in \cite{VQE2,VQE3} using a VQE with the circuit of Figure \ref{fig:circuit_vqe}. Our results in Figures \ref{fig:vqe2},\ref{fig:vqe3} indicate in general a good overlap between theoretical and computed energies, with the deviations between theoretical and computed energies probably due to the fact that the circuit of Figure \ref{fig:circuit_vqe} is not expressive enough to give better accuracies.


\subsection{Quantum Machine Learning}
\label{sec:qml}

\subsubsection{Introduction}
Linear optics has proven to be a fascinating playground for exploring the advantages offered by quantum devices over their classical counterparts. For example, as discussed in Section \ref{sec:bs}, Boson Sampling \cite{aaronson_computational_2011}, shows how a quantum device composed only of single photons, linear optical circuits, and single-photon detectors, can perform a computational task which quickly becomes unfeasible for the most powerful classical computers. 

In \cite{yee_fock_2021}, the authors show how linear optical circuits, similar to those used in Boson Sampling, can be used to solve other problems of more practical use than sampling. The basic idea is, following the work of \cite{perez2020data,schuld2021effect} for qubit-based quantum circuits, to \emph{encode} data points $x \in \R$ onto the angles of phase shifters of a universal linear optical circuit. Effectively, this allows a  \emph{non-linear}  manipulation of these data points. 
Indeed, this encoding of the data points $x$ allows to express the expectation value of some observable, computed using the linear optical circuit, as a \emph{Fourier series} 
\begin{equation} 
f(x)=\sum_{\omega \in \Omega}c_{\omega}e^{iwx} \, ,
\end{equation}  of the data points $x$ \cite{yee_fock_2021}. The Fourier series, being a well known universal (periodic) function approximator, can be used for a variety of tasks, including approximating the solution of differential equations, which we will study here.

Interestingly, $\Omega= \{ -n, \dots, n \}$  where $n$
is the number of photons inputted into the linear optical circuit \cite{yee_fock_2021}; meaning that the \emph{expressivity} of the Fourier series -- how well it can approximate a given function -- depends (among other things) on the number of input photons of the linear optical circuit.

In the coming sections, we give an example of a quantum machine learning algorithm, using the above encoding, which solves a differential equation. 

\subsubsection{Expression of Photonic Quantum Circuit Expectation Values as Fourier Series}
\label{dffeqcircuit}

We focus on constructing a universal function approximator of a one-dimensional function $f(x)$ with a photonic quantum device of $n$ photons and $m$ modes. This device consists of the following components: $n$ sources of single photons, $m$-mode \emph{universal} linear optical circuits, and $m$ number-resolving single-photon detectors \cite{hadfield2009single}. Linear optical circuits can be configured to implement $m \times m$ unitary matrices $U$. A linear optical circuit is called \emph{universal} if it can implement \emph{any} such unitary $U$ \cite{Reck94}. The $n$ single photon sources produce input Fock states of the form $|n_1,...,n_m\rangle$, where $n_i$ is the number of photons in mode $i$, and $\sum_{i=1,..,m}n_i=n$. The Fock space of $n$ photons in $m$ modes is \emph{isomorphic} to the Hilbert space $\C^{M}$, with $M=\binom{m+n-1}{n}$ \cite{aaronson_computational_2011}. This isomorphism, together with a homomorphism from $\mathsf{U}(m)$ to $\mathsf{U}(M)$ (the $m$ and $M$-dimensional unitary groups) detailed in  \cite{aaronson_computational_2011}, allows to understand the action of the $m \times m$ unitary $U$ implemented by the linear optical circuit as an $M \times M$ unitary $\mathcal{U}$ acting on the input Fock state. In the rest of this section, as in \cite{yee_fock_2021}, we will work with these $M \times M$ unitary matrices when computing the universal function approximator.

The circuit architecture of \cite{yee_fock_2021}  implements the $M \times M$ unitary transformation  \begin{equation} 
\mathcal{U}(x, \boldsymbol{\theta}):=\mathcal{W}^{(2)}\left(\boldsymbol{\theta}_{2}\right) \mathcal{S}(x) \mathcal{W}^{(1)}\left(\boldsymbol{\theta}_{1}\right) \, .
\end{equation}
The phase shift operator $\mathcal{S}(x)$ incorporates the $x$ dependency of the function we wish to approximate. It is sandwiched between two universal linear optical circuits $\mathcal{W}^{(1)}(\boldsymbol{\theta_1})$ and $\mathcal{W}^{(2)}(\boldsymbol{\theta_2})$. The parameters (angles of beam splitters and phase shifters of the linear optical circuits) $\boldsymbol{\theta_1}$ and $\boldsymbol{\theta_2}$  are tunable to enable training of the circuit, $\boldsymbol{\theta}:=\{\boldsymbol{\theta_1},\boldsymbol{\theta_2}\}$.

Let $\mathbf{\ket{n^{(i)}}} = \ket{n^{(i)}_1,n^{(i)}_2,\dots,n^{(i)}_m}$ be the input state consisting of $n$ photons where $n^{(i)}_j$ is the number of photons in input mode $j$. Consider the operator  $\mathcal{M}(\boldsymbol{\lambda})$, given by 
\begin{equation}
    \mathcal{M}(\boldsymbol{\lambda}) = \sum_{\mathbf{\ket{n^{(f)}}}}\lambda_{\mathbf{\ket{n^{(f)}}}}\mathbf{\ket{n^{(f)}}}\mathbf{\bra{n^{(f)}}} \, ,
\end{equation}
where the sum is taken over all $M$ possible Fock states $|\mathbf{n}^{(\mathbf{f})}\rangle$  of $n$ photons in $m$ modes, and $\{\boldsymbol{\lambda}_{\mathbf{\ket{n^{(f)}}}}\}$ some tunable set of parameters. The expectation value 
\begin{equation}
    f^{(n)}(x, \boldsymbol{\theta}, \boldsymbol{\lambda}):=\left\langle\mathbf{n}^{(i)}\left|\mathcal{U}^{\dagger}(x, \boldsymbol{\theta}) \mathcal{M}(\boldsymbol{\lambda}) \mathcal{U}(x, \boldsymbol{\theta})\right| \mathbf{n}^{(i)}\right\rangle \, ,
\end{equation}
of $\mathcal{M}(\boldsymbol{\lambda})$ with respect to the output state $\mathcal{U}(x, \boldsymbol{\theta}) | \mathbf{n}^{(i)} \rangle$ of the linear optical circuit  can be computed by measuring the output modes of this circuit using number-resolving detectors. 

$f^{(n)}(x, \boldsymbol{\theta}, \boldsymbol{\lambda})$ can be rewritten as the following Fourier series \cite{yee_fock_2021}
\begin{equation}
    f^{(n)}(x, \boldsymbol{\theta}, \boldsymbol{\lambda})=\sum_{\omega \in \Omega_{n}} c_{\omega}(\boldsymbol{\theta}, \boldsymbol{\lambda}) e^{i \omega x} \, ,
\end{equation}
where $\Omega_n = \llbracket -n, n \rrbracket$ is the frequency spectrum one can reach with $n$ incoming photons and $\{c_\omega(\boldsymbol{\theta}, \boldsymbol{\lambda})\}$ are the Fourier coefficients. Hence this specific architecture can be used as a universal function approximator.

\subsubsection{Application to Differential Equation Solving} The most general form of a differential equation verified by a function $f(x)$ is 
\begin{equation}
F\left[\left\{d^{m} f / d x^{m}\right\}_{m},f, x\right]=0 \, ,
\end{equation}
$F[.]$ being an operator acting on $f(x)$, its derivatives and $x$. 

Given such an expression, we wish to optimise the parameters $\boldsymbol{\theta}$ such that $f^{(n)}(x, \boldsymbol{\theta}, \boldsymbol{\lambda})$ is a good approximation to a solution $f(x)$. More precisely, in this Quantum Machine Learning task, we aim at minimising a loss function $\mathcal{L}(\boldsymbol{{\theta}})$ whose value is related to the closeness of our approximator to a solution. This is done via a classical optimisation of the quantum free parameters $\boldsymbol{\theta}$, yielding ideally in the end
\begin{equation}
    \boldsymbol{\theta^*} := \text{arg}\min_{\boldsymbol{\theta}}\mathcal{L}(\boldsymbol{{\theta}}) \, .
\end{equation}
For the solving of differential equations, the loss function described in \cite{kyriienko_solving_2021} consists of two terms
\begin{equation}
    \mathcal{L}_{\boldsymbol{\theta}}\left[\left\{d^{m} g / d x^{m}\right\}_{m},f, x\right]:=\mathcal{L}_{\boldsymbol{\theta}}^{(\mathrm{diff})}\left[\left\{d^{m} g / d x^{m}\right\}_{m},g, x\right]+\mathcal{L}_{\boldsymbol{\theta}}^{(\text {boundary})}[g, x] \, .
\end{equation}
The first term $\mathcal{L}_{\boldsymbol{\theta}}^{(\mathrm{diff})}\left[\left\{d^{m} g / d x^{m}\right\}_{m},g, x\right]$ corresponds to the differential equation which has been discretised over a set of $M$ points $\{x_i\}$:
\begin{equation}
    \mathcal{L}_{\boldsymbol{\theta}}^{(\mathrm{diff})}\left[\left\{d^{m} g / d x^{m}\right\}_{m},g, x\right]:=\frac{1}{M} \sum_{i=1}^{M} L\left(F\left[d_{x}^m g\left(x_{i}\right), g\left(x_{i}\right), x_{i}\right], 0\right) \, ,
\end{equation}
where $L(a,b) := (a - b)^2$ for $a,b \in \R$. The second term $\mathcal{L}_{\boldsymbol{\theta}}^{(\text{boundary})}[g, x]$ is associated to the initial conditions of our desired solution. It is defined as 
\begin{equation}
    \mathcal{L}_{\boldsymbol{\theta}}^{\text {(boundary) }}[g, x]:=\eta L\left(g(x_0), f_{0}\right) \, ,
\end{equation}
where $\eta$ is the weight granted to the boundary condition and $f_{0}$ is given by $f(x_0) = f_0$, for some initial data point $x_0$. These functions will be applied to the iterated approximations $f^{(n)}(x, \boldsymbol{\theta}, \boldsymbol{\lambda})$.

\subsubsection{\perceval Implementation\texorpdfstring{\protect\footnotemark}{} }
\footnotetext{This implementation is accompanied by a \href{https://perceval.quandela.net/docs/notebooks/Differential\%20equation\%20solving.html}{notebook}.}

Our aim is to reproduce some of the results of \cite{kyriienko_solving_2021}, where the authors use so-called differentiable quantum circuits, together with the classical optimisation BFGS method of \href{https://scipy.org/}{SciPy} \cite{2020SciPyNMeth}, to provide an approximation to the solution of the nonlinear differential equation 
\begin{equation} 
\frac{d f}{d x}+\lambda f(x)(\kappa+\tan (\lambda x))=0,
\end{equation}  with  $\lambda, \kappa \in \R$, and boundary condition $f(0)=f_{0} \in \R$. The analytical solution  of this differential equation is \begin{equation} 
f(x)=\exp (-\kappa \lambda x) \cos (\lambda x)+ f_0 - 1.
\end{equation}  
Here, we solve this differential equation using the linear optical circuit architectures of Section \ref{dffeqcircuit} simulated in \perceval, together with classical optimisation. Note 
that the authors of \cite{kyriienko_solving_2021} use Chebyshev polynomials as  universal function approximators, whereas here we will use the Fourier series.

In our \perceval implementation, $\eta$ is chosen empirically as $\eta = 5$, granting sufficient weight to the boundary condition. Concerning the $\boldsymbol{\lambda}$ parameters, each one of them is sampled uniformly randomly in the interval $\{-200, \dots, 200\}$, which is also empirically chosen. One could tune these $\boldsymbol{\lambda}$ parameters as well for greater accuracy, at the cost of a considerably slower minimisation procedure. The number of modes $m$ is taken to be equal to the number of photons $n$. Differentiation is numerically conducted, $\frac{df}{dx} \simeq \frac{\Delta f}{\Delta x}$, $\Delta x$ taken as $10^{-3}$. The discretised version of the loss function is taken on a grid of $50$ points, uniformly spaced between $0$ and $1$.

\begin{figure}[ht]
	\centering
	\begin{subfigure}[t]{0.45\textwidth}
		\centering
		\includegraphics[width=\textwidth]{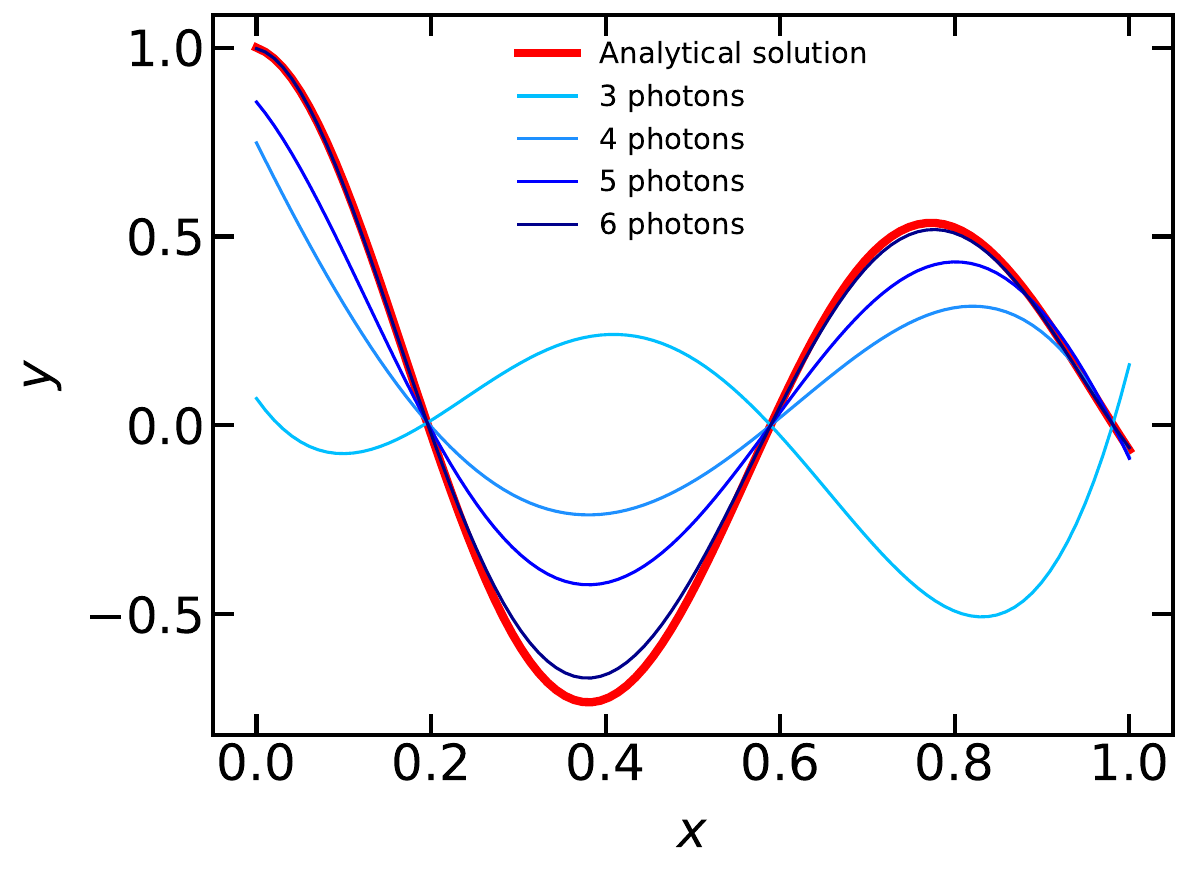}
		\caption{Comparison of the analytical solution to the considered differential equation to converged solutions of the discretised quantum loss function in terms of input photon number. Parameters of the differential equation are taken as $\lambda=8$, $\kappa=0.1$, $f_0=1$ matching that of \cite{kyriienko_solving_2021}.}
		\label{fig:diff_equation_approximation}
	\end{subfigure}
	\quad
	\begin{subfigure}[t]{0.45\textwidth}
		\centering
		\includegraphics[width=\textwidth]{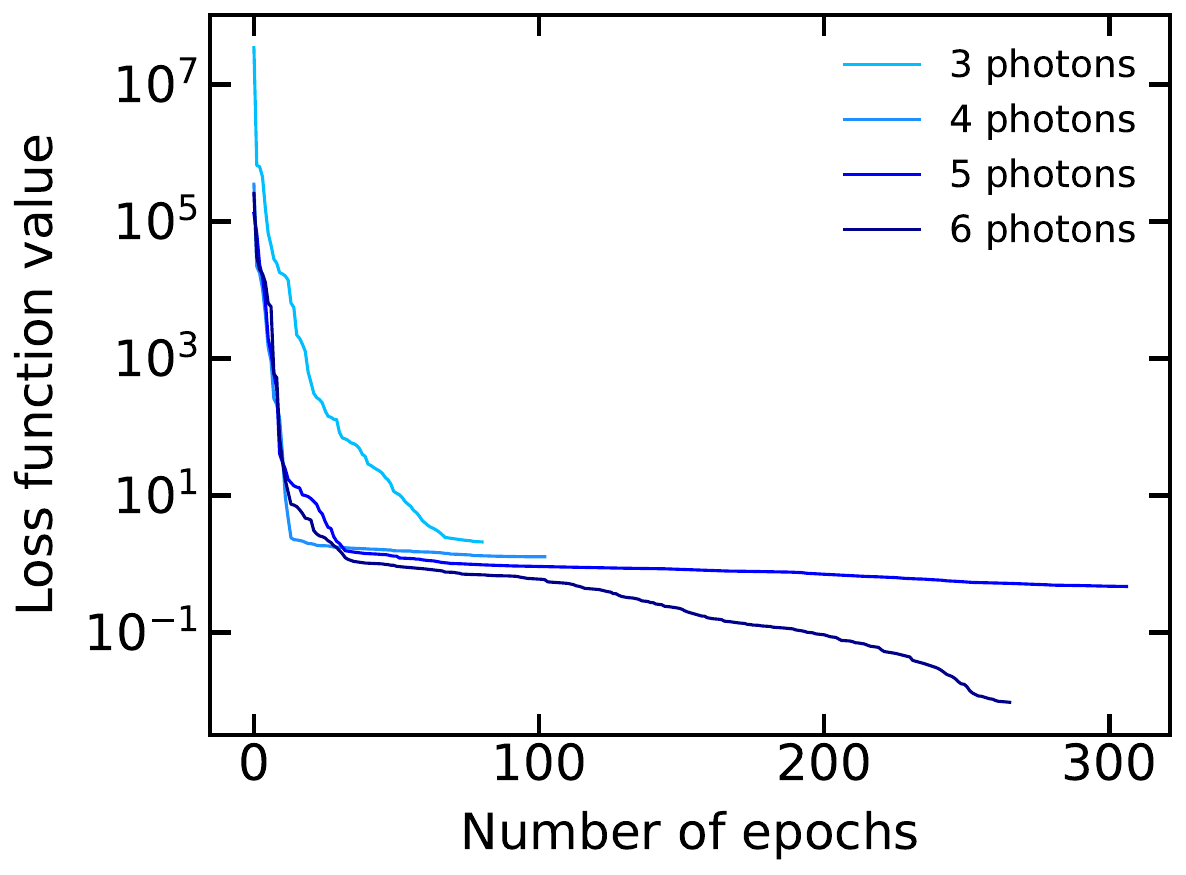}
		\caption{Loss evolution as a function of the number of epochs for various input photon numbers.}
		\label{fig:loss_evolution}
	\end{subfigure}
  \caption{Results of QML simulation using \perceval.}
	\label{fig:qml_res}
\end{figure}

Results are shown for various photon numbers in Figure \ref{fig:diff_equation_approximation}, demonstrating the increase in expressivity of the quantum circuit with growing $n$. 
The converged solution for $6$ input photons provides a convincing approximation to the analytical solution. Increasing the number of photons further should yield even more accurate results, allowing for the realisation of more complex Quantum Machine Learning tasks. 

A common indicator of the performance of a machine learning algorithm is the convergence of its loss function \cite{raschka2019python}.
Confirming the results from Figure \ref{fig:diff_equation_approximation}, additional photons allow us to reach lower values of the loss function, as shown in Figure \ref{fig:loss_evolution}. However this results in a longer convergence time due to an increased number of tunable parameters. Further details concerning the simulation in \perceval can be found in the  \href{https://github.com/Quandela/Perceval}{documentation}.

\bigskip
To summarise, we have used \perceval to simulate linear optical circuits providing universal function approximators \cite{yee_fock_2021}, and shown that these can be used together with techniques from Quantum Machine Learning to accurately compute the solutions of differential equations. The accuracy of these function approximators depends, among other things, on the number of input photons of the linear optical circuits.
An interesting future direction we aim to pursue is using \perceval and our developed techniques to solve other types of differential equations of significant practical interest \cite{widder1976heat,constantin2020navier}.

%% file: s5_conclusion.tex
\section{Conclusion}
\perceval is a unique framework dedicated to linear optics and photonic quantum computing.
This white paper has aimed to provide an overview of the platform, the motivations for its development, its structure and main features, and to give a variety of examples of \perceval in action.
These examples are intended to be illustrative of some of the immediate uses of the platform.

\perceval's simulation back-ends are optimised to run on local desktop devices, with extensions for HPC clusters.
They can be used to run computational experiments to fine-tune algorithms, compare with experimental data from actual experiments and photonic quantum computing platforms, and can reproduce published articles in few lines of code.

\perceval is intended to be accessible to physicists, both experimental and theoretical, and computer scientists alike, with a goal of providing a bridge between these communities.
With the intention of keeping a strong connection between software and hardware for photonic quantum computing, a major focus of future development will be on the continued development of realistic noise-models, that can describe with increasing accuracy the functioning of specific hardware components.

\perceval allows users to design algorithms and linear optical circuits through a large collection of predefined components.
The collection of algorithms described here are available and presented as tutorials in the \href{https://perceval.quandela.net/docs/index.html}{documentation}.
This is an open source project, with a modular architecture, and is welcoming of contributions from the community.

It is intended that future versions of \perceval will include more optimised simulators, noisy simulators, features for working with density matrices and cluster states, more advanced features and options for detectors to cover both threshold and photon-number resolving detectors, as well as features for treating circuits with feedforward.

%% file: appendix.tex

\newcommand{\backcap}[1]{Example of how to use the back-end {\tt #1} described in Section \ref{subsub:back}}

\newcommand{\codeblock}[3]{
\begin{lstlisting}[language=python,style=mypython,caption= #1,captionpos=b,label=#2]^^J
#3^^J
\end{lstlisting}
}

\newcommand{\backfig}[2]{
\begin{figure}[h]
	\centering
	\includegraphics{#1}
    \caption{Output of the Code \ref{#2}.}
	\label{fig:#2}
\end{figure}
}

\section{Examples Codes of the Back-ends and the {\tt Processor} class}

\subsection{{\tt CliffordClifford2017}}

\begin{lstlisting}[language=python,style=mypython,caption=Example of how to use the back-end {\tt CliffordClifford2017} described in Section \ref{subsub:back},captionpos=b,label=code:CliffordClifford2017]
import perceval as pcvl
from perceval.components import Unitary

circuit = Unitary(pcvl.Matrix.random_unitary(8))
input_state = pcvl.BasicState([1, 0]*4)
nsamples = 5
clifford_backend = pcvl.BackendFactory.get_backend("CliffordClifford2017")
circuit_simulator = clifford_backend(circuit)
# Clifford&Clifford back-end is specialized in sampling tasks
samples = circuit_simulator.samples(input_state, nsamples)
for s in samples:
    print(s)
\end{lstlisting}

\begin{figure}[ht]
    \centering
    \texttt{|1,0,0,0,0,1,2,0> \\
|0,1,1,1,1,0,0,0> \\
|0,0,4,0,0,0,0,0> \\
|1,0,0,0,0,3,0,0> \\
|1,0,0,0,1,1,1,0>}\hfill
    \caption{Output of Code \ref{code:CliffordClifford2017}}
    \label{fig:cc_output}
\end{figure}

\subsection{{\tt Naive}}

\begin{lstlisting}[language=python,style=mypython,caption=Example of how to use the back-end {\tt Naive} described in Section \ref{subsub:back},captionpos=b,label=code:Naive]
import perceval as pcvl
from perceval.components import Unitary

circuit = Unitary(pcvl.Matrix.random_unitary(4))
input_state = pcvl.BasicState([1, 0, 1, 0])
output_states = [pcvl.BasicState([1, 0, 1, 0]),
                 pcvl.BasicState([1, 0, 0, 1]),
                 pcvl.BasicState([0, 1, 1, 0]),
                 pcvl.BasicState([0, 0, 2, 0])]
nsamples = 5
naive_backend = pcvl.BackendFactory.get_backend("Naive")
circuit_simulator = naive_backend(circuit)
# Naive is able to compute probabilities and probability amplitudes
# for an input / output state pair
for os in output_states:
    p = circuit_simulator.prob(input_state, os)
    pa = circuit_simulator.probampli(input_state, os)
    print(f"{input_state} -> {os}: p={p}, p.ampl.={pa}")
\end{lstlisting}

\begin{figure}[ht]
    \centering
    \texttt{|1,0,1,0> -> |1,0,1,0>: p=0.11351, p.ampl.=(-0.23806+0.23841j) \\|1,0,1,0> -> |1,0,0,1>: p=0.09711, p.ampl.=(0.30757+0.05013j) \\|1,0,1,0> -> |0,1,1,0>: p=0.19920, p.ampl.=(-0.02982+0.44532j) \\|1,0,1,0> -> |0,0,2,0>: p=0.11104, p.ampl.=(-0.33287-0.01555j)}\hfill
    \caption{Output of Code \ref{code:Naive}}
    \label{fig:naive_output}
\end{figure}

\subsection{{\tt SLOS}}

\begin{lstlisting}[language=python,style=mypython,caption=Example of how to use the back-end {\tt SLOS} described in Section \ref{subsub:back},captionpos=b,label=code:slos]
import perceval as pcvl
from perceval.components import Unitary

circuit = Unitary(pcvl.Matrix.random_unitary(4))
input_state = pcvl.BasicState([1, 0, 1, 0])
nsamples = 5
slos_backend = pcvl.BackendFactory.get_backend("SLOS")
circuit_simulator = slos_backend(circuit)
# SLOS computes all output probabilities at once
# They can be retrieved via a specialized iterator in the back-end
state_distribution = pcvl.BSDistribution()
for ostate, prob in circuit_simulator.allstateprob_iterator(input_state):
    state_distribution[ostate] = prob
pcvl.pdisplay(state_distribution, output_format=pcvl.Format.LATEX)
\end{lstlisting}

\begin{figure}[ht]
    \centering
    {\tt \begin{tabular}{lr}
    \hline
    state     &   probability \\
    \hline
    |1,0,1,0\ensuremath{>} &   0.18959     \\
    |2,0,0,0\ensuremath{>} &   0.162852    \\
    |0,0,1,1\ensuremath{>} &   0.158684    \\
    |0,0,2,0\ensuremath{>} &   0.138282    \\
    |0,0,0,2\ensuremath{>} &   0.117126    \\
    |1,0,0,1\ensuremath{>} &   0.088445    \\
    |0,1,1,0\ensuremath{>} &   0.075928    \\
    |0,1,0,1\ensuremath{>} &   0.056393    \\
    |0,2,0,0\ensuremath{>} &   0.01188     \\
    |1,1,0,0\ensuremath{>} &   0.000819328 \\
    \hline
    \end{tabular}}
    \caption{Output of Code \ref{code:slos}}
    \label{fig:slos_output}
\end{figure}

\subsection{{\tt Processor} and {\tt Algorithm}}
\label{app:proc}
\begin{lstlisting}[language=python,style=mypython,caption=Example of how to use the {\tt Processor} described in Section \ref{subsub:processors},captionpos=b,label=code:processors]
import perceval as pcvl
import numpy as np
from perceval.algorithm import Sampler, Analyzer
from perceval.components import Processor, Source

# SLOS backend does not support sampling natively
# However, the Sampler algorithm is able to reconstruct sampling results, transparently, through probability computing
cnot = pcvl.Circuit(6, "Raplh CNOT")
cnot.add((3, 4), pcvl.BS())
cnot.add((0, 1), pcvl.BS(pcvl.BS.r_to_theta(1 / 3), phi_bl=np.pi, phi_tr=np.pi/2, phi_tl=-np.pi/2))
cnot.add((2, 3), pcvl.BS(pcvl.BS.r_to_theta(1 / 3), phi_bl=np.pi, phi_tr=np.pi/2, phi_tl=-np.pi/2))
cnot.add((4, 5), pcvl.BS(pcvl.BS.r_to_theta(1 / 3)))
cnot.add((3, 4), pcvl.BS())

def postselection_func(ostate):
    return ostate[0] == 0 and (ostate[1] + ostate[2] == 1) \
        and ostate[5] == 0 and (ostate[3] + ostate[4] == 1)

cnot_processor = pcvl.Processor("SLOS", cnot)
cnot_processor.set_postprocess(postselection_func)

nsample = 50000
sampler = Sampler(cnot_processor)
cnot_processor.with_input(pcvl.BasicState([0, 1, 0, 1, 0, 0]))  # Corresponds to logical qubit state |0,0>
output = sampler.sample_count(nsample)
pcvl.pdisplay(output['results'], output_format=pcvl.Format.LATEX)
print(f"Ratio of samples with 2 photons: {pcvl.simple_float(output['physical_perf'])[0]}")
print(f"Gate performance: {pcvl.simple_float(output['logical_perf'])[0]}")

cnot_processor.source = Source(emission_probability=0.5)  # Now use a source with a 50% first lens brigthness
# With an imperfect source, the expected state is turned to an actual input distribution by the source model.
cnot_processor.with_input(pcvl.BasicState([0, 1, 0, 1, 0, 0]))
output = sampler.sample_count(nsample)
# Here we expect perfect results, with a low physical performance
# (a lot of samples are discarded because they do not produce a 2 photon coincidence)
pcvl.pdisplay(output['results'], output_format=pcvl.Format.LATEX)
print(f"Ratio of samples with 2 photons: {pcvl.simple_float(output['physical_perf'])[0]}")
print(f"Gate performance: {pcvl.simple_float(output['logical_perf'])[0]}")

# Now use a source that may emit distinguishable photons, or 2 photons at once
cnot_processor.source = Source(emission_probability=0.5,
                               indistinguishability=0.95,
                               multiphoton_component=0.1)
cnot_processor.with_input(pcvl.BasicState([0, 1, 0, 1, 0, 0]))
output = sampler.sample_count(nsample)
# Here the results start being noisy
pcvl.pdisplay(output['results'], output_format=pcvl.Format.LATEX)
print(f"Ratio of samples with 2 photons: {pcvl.simple_float(output['physical_perf'])[0]}")
print(f"Gate performance: {pcvl.simple_float(output['logical_perf'])[0]}")
\end{lstlisting}

\begin{figure}[ht]
    \centering
    {\tt \begin{tabular}{lr}
    \hline
     state         &   count \\
    \hline
     |0,1,0,1,0,0\ensuremath{>} &   50000 \\
    \hline
    \end{tabular} \\ \vspace{0.1cm}
    Ratio of samples with 2 photons: 1 \\ \vspace{0.1cm}
    Gate performance: 1/9 \\ \vspace{0.1cm}
    \begin{tabular}{lr}
    \hline
     state         &   count \\
    \hline
     |0,1,0,1,0,0\ensuremath{>} &   50000 \\
    \hline
    \end{tabular}  \\ \vspace{0.1cm}
    Ratio of samples with 2 photons: 1/4  \\ \vspace{0.1cm}
    Gate performance: 1/9 \\ \vspace{0.1cm}
    \begin{tabular}{lr}
    \hline
     state         &   count \\
    \hline
     |0,1,0,1,0,0\ensuremath{>} &   47347 \\
     |0,0,1,1,0,0\ensuremath{>} &    2653 \\
    \hline
    \end{tabular} \\\vspace{0.1cm}
    Ratio of samples with 2 photons: 0.257315 \\\vspace{0.1cm}
    Gate performance: 0.108027}
    \caption{Output of Code \ref{code:processors}}
    \label{fig:slos_output2}
\end{figure}

\newpage

\newcommand{\actioncap}[1]{Code implementing the Section #1}

\section{Codes of Section \ref{sec:action}}
\label{app:examples}

\subsection{The Hong-Ou-Mandel Effect}

\begin{lstlisting}[language=python,style=mypython,caption=Example code - \ref{sec:hom} Hong-Ou-Mandel Effect,captionpos=b,label=code:appHOM]
import perceval as pcvl
from random import random
from collections import Counter
import matplotlib.pyplot as plt

# register_click function will register each consecutive click
# and calculate the distance between
# click on different arms of the inteferometer

# last_click is where we found the last photon
last_click = None
# distance is how many click away it was
distance = None

def register_click(sample):
    global counts, last_click, distance
    if sample[0] and sample[1]:
        counts[0] += 1
        last_click = None
    else:
        if sample[1]:
            if last_click == 0:
                counts[distance] += 1
            last_click = 1
            distance = 1
        elif sample[0]:
            if last_click == 1:
                counts[-distance] += 1
            last_click = 0
            distance = 1
        elif distance:
            distance += 1

# imperfect source with a slight g_2
source = pcvl.Source(multiphoton_component=0.01)

# we define the circuit
HOM = pcvl.Processor("SLOS", 2, source)
HOM.add((0,1), pcvl.BS())
HOM.add(1, pcvl.TD(1))
HOM.add((0,1), pcvl.BS())
pcvl.pdisplay(HOM)

# the circuit is "expanded" - ie, time loop is converted into
# additional modes
# and we define a new processor for this circuit 
expand_components, extend_m = pcvl.computation.expand_TD(HOM.flatten(), 2, 2, 1, False)
p = pcvl.Processor("CliffordClifford2017", extend_m, source)
for r, c in expand_components:
    p.add(r, c)
pcvl.pdisplay(p)

counts = Counter()
photon_delay = 0
# generate 2000 photons, since we have expanded time in
# additional modes, each iteration is 2 time-steps
for i in range(1000):
    # photon have 30% of being emitted
    input = [random()>0.3 and 1 or 0, 0, random()>0.3 and 1 or 0, 0, photon_delay]
    if sum(input):
        # if there is at least one photon
        p.with_input(pcvl.BasicState(input))
        out = pcvl.algorithm.Sampler(p).samples(1)["results"][0]
        register_click(out[0:2])
        register_click(out[2:4])
        photon_delay=out[4]
    else:
        # otherwise, we register the 2 following clicks as 0
        register_click((0,0))
        register_click((0,0))
        photon_delay=0

# the distribution
print(counts)

fig = plt.figure()
dist = list(range(-2,3))
ax = fig.add_axes([0,0,1,1])
count_dist = [counts[d] for d in dist]
ax.bar(dist, count_dist)
plt.show()
\end{lstlisting}

\subsection{Boson Sampling}

\begin{lstlisting}[language=python,style=mypython,caption=Example code - \ref{sec:bs} Boson Sampling,captionpos=b,label=code:appBosonsSampling]
from collections import Counter
import gzip
import pickle
import time
import random
import perceval as pcvl
from perceval.algorithm import Sampler

n = 14       #number of photons at the input
m = 60       #number of modes
N = 5000000  #number of samples
Unitary_60 = pcvl.Matrix.random_unitary(m)

mzi = (pcvl.BS() // (0, pcvl.PS(phi=pcvl.Parameter("\phi_a")))
       // pcvl.BS() // (1, pcvl.PS(phi=pcvl.Parameter("\phi_b"))))
Linear_Circuit_60 = pcvl.Circuit.decomposition(Unitary_60, mzi,
                                               phase_shifter_fn=pcvl.PS,
                                               shape="triangle")
QPU = pcvl.Processor("CliffordClifford2017", Linear_Circuit_60)

#one can choose which mode he/she wants at input, or we can choose it randomly
def Generating_Input(n, m, modes = None):
    "This function randomly chooses an input with n photons in m modes."
    if modes == None :
        modes = sorted(random.sample(range(m),n))
    state = "|"
    for i in range(m):
        state = state + "0"*(1 - (i in modes)) +"1"*(i in modes)+ ","*(i < m-1)
    return pcvl.BasicState(state + ">")

input_state = Generating_Input(n, m)
QPU.with_input(input_state)

# Keep all outputs
QPU.mode_post_selection(0)

sampler = Sampler(QPU)

# if we want to launch parallel process
worker_id=1

#store the input and the unitary
with open("%dphotons_%dmodes_%dsamples-worker%s-unitary.pkl" %(n,m,N,worker_id), 'wb') as f:
    pickle.dump(Unitary_60, f)

with open("%dphotons_%dmodes_%dsamples-worker%s-inputstate.pkl" %(n,m,N,worker_id), 'w') as f:
    f.write(str(input_state)+"\n")

with gzip.open("%dphotons_%dmodes_%dsamples-worker%s-samples.txt.gz" %(n,m,N,worker_id), 'wb') as f:
    start = time.time()
    for _ in range(N):
        f.write((str(sampler.samples(1)["results"][0])+"\n").encode())
    end = time.time()
    f.write(str("==> %d\n" % (end-start)).encode())

count = 0
bunching_distribution = Counter()

with gzip.open("%dphotons_%dmodes_%dsamples-worker%s-samples.txt.gz"%(n,m,N,worker_id), "rt") as f:
    for l in f:
        l = l.strip()
        if l.startswith("|") and l.endswith(">"):
            try:
                st = pcvl.BasicState(l)
                count+=1
                bunching_distribution[st.photon2mode(st.n-1)]+=1
            except Exception:
                pass
print(count, "samples")
print("Bunching Distribution:", "\t".join([str(bunching_distribution[k]) for k in range(m)]))
\end{lstlisting}

\subsection{Grover's Algorithm}

\begin{lstlisting}[language=python,style=mypython,caption=Example code - \ref{subsec:grover} Grover's algorithm,captionpos=b,label=code:appGrover]
import numpy as np
import sympy as sp
import matplotlib.pyplot as plt
import perceval as pcvl

states = [pcvl.BasicState("|0,{P:H}>"),
          pcvl.BasicState("|0,{P:V}>"),
          pcvl.BasicState("|{P:H},0>"),
          pcvl.BasicState("|{P:V},0>"),
         ]

states_modes = [
    pcvl.BasicState([0, 0, 0, 1]),
    pcvl.BasicState([0, 0, 1, 0]),
    pcvl.BasicState([0, 1, 0, 0]),
    pcvl.BasicState([1, 0, 0, 0])
]

BS = pcvl.BS.Ry()
pcvl.pdisplay(BS.U)

def HWP(xsi):
    hwp = pcvl.Circuit(m=1)
    hwp.add(0, pcvl.HWP(xsi)).add(0, pcvl.PS(-sp.pi/2))
    return hwp

init_circuit = pcvl.Circuit(m=2, name="Initialization")
init_circuit.add(0, HWP(sp.pi/8))
init_circuit.add((0, 1), BS)
init_circuit.add(0, pcvl.PS(-sp.pi))

def oracle(mark):
    """Values 0, 1, 2 and 3 for parameter 'mark' respectively mark the elements "00", "01", "10" and "11" of the list."""
    oracle_circuit = pcvl.Circuit(m=2, name='Oracle')
    # The following dictionnary translates n into the corresponding component settings
    oracle_dict = {0: (1, 0), 1: (0, 1), 2: (1, 1), 3: (0, 0)}
    PC_state, LC_state = oracle_dict[mark]
    # Mode b
    if PC_state == 1:
        oracle_circuit.add(0, HWP(0))
    oracle_circuit.add(0, pcvl.PR(sp.pi/2))
    if LC_state == 1:
        oracle_circuit.add(0, HWP(0))
    # Mode a
    if LC_state == 1:
        oracle_circuit.add(1, HWP(0))
    if PC_state == 1:
        oracle_circuit.add(1, HWP(0))
    return oracle_circuit

inversion_circuit = pcvl.Circuit(m=2, name='Inversion')
inversion_circuit.add((0, 1), BS)
inversion_circuit.add(0, HWP(sp.pi/4))
inversion_circuit.add((0, 1), BS)

detection_circuit = pcvl.Circuit(m=4, name='Detection')
detection_circuit.add((0, 1), pcvl.PBS())
detection_circuit.add((2, 3), pcvl.PBS())

def grover_circuit(mark):
    grover_circuit = pcvl.Circuit(m=4, name='Grover')
    grover_circuit.add(0, init_circuit).add(0, oracle(mark)).add(0, inversion_circuit)
    grover_circuit.add(1, pcvl.PERM([1, 0])).add(0, detection_circuit)
    return grover_circuit

# Circuit simulation
input_state = pcvl.BasicState("|{P:H},0, 0, 0>")
results_list = []  # probability amplitudes storage

for mark in range(4):
    sim = pcvl.Processor("Naive", grover_circuit(mark))
    ca = pcvl.algorithm.Analyzer(sim,
                                 input_states=[input_state],
                                 output_states=states_modes,
                                 )
    results_list.append(ca.distribution[0])

# Plot data
labels = ['"00"', '"01"', '"10"', '"11"']
state_0_prob_list = results_list[0]
state_1_prob_list = results_list[1]
state_2_prob_list = results_list[2]
state_3_prob_list = results_list[3]
x = np.arange(4)  # label locations
width = 0.1  # the width of the bars

fig, ax = plt.subplots(dpi=150)
rects_0 = ax.bar(x - 3 * width / 2, state_0_prob_list, width, label=str(states[0]))
rects_1 = ax.bar(x - width / 2, state_1_prob_list, width, label=str(states[1]))
rects_2 = ax.bar(x + width / 2, state_2_prob_list, width, label=str(states[2]))
rects_3 = ax.bar(x + 3 * width / 2, state_3_prob_list, width, label=str(states[3]))

ax.set_xlabel('Marked database element')
ax.set_ylabel('Detection probability')
ax.set_xticks(x, labels)
ax.legend()
ax.grid(True, axis='x')
plt.show()
\end{lstlisting}

\subsection{Shor's Algorithm}

\begin{lstlisting}[language=python,style=mypython,caption=Example code - \ref{sec:shor} Shor's algorithm,captionpos=b,label=code:appShor]
import perceval as pcvl

def toFockState(qubitState):
    # path encoding
    pe = {0:[1,0],  1:[0,1]}
    return [0] + pe[qubitState[0]] \
        + pe[qubitState[2]] + [0, 0] \
        + pe[qubitState[1]] + pe[qubitState[3]] + [0]

def toQubitState(fockState):
    # qubit modes
    x1 = [1, 2]
    f1 = [3, 4]
    x2 = [7, 8]
    f2 = [9, 10]
    # auxiliary modes
    am1 = [0, 5]
    am2 = [6, 11]

    # auxiliary modes
    for i in am1 + am2:
        if fockState[i] != 0:
            return None
    L = []
    # qubit modes
    for q in [x1, x2, f1, f2]:
        if fockState[q[0]] + fockState[q[1]] != 1:
            return None
        else:
            L.append(fockState[q[1]])
    return L

def strState(state):
    return str(pcvl.BasicState(state))

# Build the circuit
circ = pcvl.Circuit(12)

# qubit modes
# for qubit states 0, 1
x1 = [1, 2]
f1 = [3, 4]
x2 = [7, 8]
f2 = [9, 10]
# auxiliary modes
am1 = [0, 5]
am2 = [6, 11]

# H gates
for q in [x1, f1, x2, f2]:
    circ.add(q, pcvl.BS.H())

# CZ gates
for x, f, am in [(x1, f1, am1), (x2, f2, am2)]:
    circ.add((am[0], x[0]), pcvl.BS(pcvl.BS.r_to_theta(1/3))) # R = 1/3
    circ.add((x[1],  f[0]), pcvl.BS(pcvl.BS.r_to_theta(1/3)))
    circ.add((f[1], am[1]), pcvl.BS(pcvl.BS.r_to_theta(1/3)))

# H gates
for q in [f1, f2]:
    circ.add(q, pcvl.BS.H())

# Create input state
qubit_istate = [0,0,0,1]
istate = toFockState(qubit_istate)

# Simulation
backend = pcvl.BackendFactory().get_backend("Naive")
simulator = backend(circ)

output_qubit_states = [
    [x1,x2,f1,f2]
    for x1 in [0,1] for x2 in [0,1] for f1 in [0,1] for f2 in [0,1]
]

print("Output state amplitudes: (post-selected on qubit states, not renormalized)")
print("|x1,x2,f1,f2>")
for oqstate in output_qubit_states:
    ostate = toFockState(oqstate)
    a = simulator.probampli(pcvl.BasicState(istate), pcvl.BasicState(ostate))
    print(strState(oqstate), a)

input_states = {
    pcvl.BasicState(pcvl.BasicState(istate)): strState(qubit_istate)
}

expected_output_states = {
    pcvl.BasicState(toFockState([x1,x2,x1,1-x2])): strState([x1,x2,x1,1-x2])
    for x1 in [0,1] for x2 in [0,1]
}

p = pcvl.Processor("Naive", circ)

ca = pcvl.algorithm.Analyzer(p, input_states, expected_output_states)
ca.compute()

print("Output state distribution: (post-selected on expected qubit states, not renormalized)")
print("|x1,x2,f1,f2>")
pcvl.pdisplay(ca)
\end{lstlisting}

\subsection{Variational Quantum Eigensolver}

\begin{lstlisting}[language=python,style=mypython,caption=Example code - \ref{sec:vqe} Variational Quantum Eigensolver,captionpos=b,label=code:appVQE]
from tqdm.auto import tqdm
import numpy as np
from scipy.optimize import minimize
import random
import matplotlib.pyplot as plt
import perceval as pcvl

simulator_backend = pcvl.BackendFactory().get_backend("Naive")

#List of the parameters phi1,phi2,...,phi8
List_Parameters=[]

# VQE is a 6 optical mode circuit
VQE=pcvl.Circuit(6)

VQE.add((1,2), pcvl.BS())
VQE.add((3,4), pcvl.BS())
List_Parameters.append(pcvl.Parameter("phi1"))
VQE.add((2,),pcvl.PS(phi=List_Parameters[-1]))
List_Parameters.append(pcvl.Parameter("phi3"))
VQE.add((4,),pcvl.PS(phi=List_Parameters[-1]))
VQE.add((1,2), pcvl.BS())
VQE.add((3,4), pcvl.BS())
List_Parameters.append(pcvl.Parameter("phi2"))
VQE.add((2,),pcvl.PS(phi=List_Parameters[-1]))
List_Parameters.append(pcvl.Parameter("phi4"))
VQE.add((4,),pcvl.PS(phi=List_Parameters[-1]))

# CNOT ( Post-selected with a success probability of 1/9)
VQE.add([0,1,2,3,4,5], pcvl.PERM([0,1,2,3,4,5]))#Identity PERM (permutation) for the purpose of drawing a nice circuit
VQE.add((3,4), pcvl.BS())
VQE.add([0,1,2,3,4,5], pcvl.PERM([0,1,2,3,4,5]))#Identity PERM (permutation) for the same purpose
VQE.add((0,1), pcvl.BS(pcvl.BS.r_to_theta(1/3)))
VQE.add((2,3), pcvl.BS(pcvl.BS.r_to_theta(1/3)))
VQE.add((4,5), pcvl.BS(pcvl.BS.r_to_theta(1/3)))
VQE.add([0,1,2,3,4,5], pcvl.PERM([0,1,2,3,4,5]))#Identity PERM (permutation) for the same purpose
VQE.add((3,4), pcvl.BS())
VQE.add([0,1,2,3,4,5], pcvl.PERM([0,1,2,3,4,5]))#Identity PERM (permutation) for the same purpose

List_Parameters.append(pcvl.Parameter("phi5"))
VQE.add((2,),pcvl.PS(phi=List_Parameters[-1]))
List_Parameters.append(pcvl.Parameter("phi7"))
VQE.add((4,),pcvl.PS(phi=List_Parameters[-1]))
VQE.add((1,2), pcvl.BS())
VQE.add((3,4), pcvl.BS())
List_Parameters.append(pcvl.Parameter("phi6"))
VQE.add((2,),pcvl.PS(phi=List_Parameters[-1]))
List_Parameters.append(pcvl.Parameter("phi8"))
VQE.add((4,),pcvl.PS(phi=List_Parameters[-1]))
VQE.add((1,2), pcvl.BS())
VQE.add((3,4), pcvl.BS())

# Mode 0 and 5 are auxillary.
#1st qubit is path encoded in modes 1 & 2
#2nd qubit in 3 & 4

#Input states of the photonic circuit
input_states = {
    pcvl.BasicState([0,1,0,1,0,0]):"|00>"}

#Outputs in the computational basis
output_states = {
    pcvl.BasicState([0,1,0,1,0,0]):"|00>",
    pcvl.BasicState([0,1,0,0,1,0]):"|01>",
    pcvl.BasicState([0,0,1,1,0,0]):"|10>",
    pcvl.BasicState([0,0,1,0,1,0]):"|11>"}

def minimize_loss(lp=None):
    # Updating the parameters on the chip
    for idx, p in enumerate(lp):
        List_Parameters[idx].set_value(p)

    # Simulation, Quantum processing part of the VQE
    s_VQE = simulator_backend(VQE.compute_unitary(use_symbolic=False))

    # Collecting the output state of the circuit
    psi = []
    for input_state in input_states:
        for output_state in output_states:  # |00>,|01>,|10>,|11>
            psi.append(s_VQE.probampli(input_state, output_state))

        # Evaluating the mean value of the Hamiltonian.  # The Hamiltonians H is defined in the following block
    psi_prime = np.dot(H[R][1], psi)
    loss = np.real(sum(sum(np.conjugate(psi) * np.array(psi_prime[0])))) / (sum([i * np.conjugate(i) for i in psi]))
    loss = np.real(loss)

    tq.set_description('%g / %g  loss function=%g' % (R, len(H), loss))
    return (loss)

# Hamiltonian #1
Hamiltonian_elem = np.array([[[0,0,0,0],[0,0,0,0],[0,0,0,0],[0,0,0,0]],    #00
                             [[1,0,0,0],[0,1,0,0],[0,0,1,0],[0,0,0,1]],    #II
                             [[0,1,0,0],[1,0,0,0],[0,0,0,1],[0,0,1,0]],    #IX
                             [[1,0,0,0],[0,-1,0,0],[0,0,1,0],[0,0,0,-1]],  #IZ
                             [[0,0,1,0],[0,0,0,1],[1,0,0,0],[0,1,0,0]],    #XI
                             [[0,0,0,1],[0,0,1,0],[0,1,0,0],[1,0,0,0]],    #XX
                             [[0,0,1,0],[0,0,0,-1],[1,0,0,0],[0,-1,0,0]],  #XZ
                             [[1,0,0,0],[0,1,0,0],[0,0,-1,0],[0,0,0,-1]],  #ZI
                             [[0,1,0,0],[1,0,0,0],[0,0,0,-1],[0,0,-1,0]],  #ZX
                             [[1,0,0,0],[0,-1,0,0],[0,0,-1,0],[0,0,0,1]]]) #ZZ

Hamiltonian_coef = np.matrix(
# [R,II,IX,IZ,XI,XZ,XX,ZI,ZX,ZZ]
[[0.05,33.9557,-0.1515,-2.4784,-0.1515,0.1412,0.1515,-2.4784,0.1515,0.2746],
[0.1,13.3605,-0.1626,-2.4368,-0.1626,0.2097,0.1626,-2.4368,0.1626,0.2081],
[0.15,6.8232,-0.1537,-2.3801,-0.1537,0.2680,0.1537,-2.3801,0.1537,0.1512],
[0.2,3.6330,-0.1405,-2.2899,-0.1405,0.3027,0.1405,-2.2899,0.1405,0.1176],
[0.25,1.7012,-0.1324,-2.1683,-0.1324,0.3211,0.1324,-2.1683,0.1324,0.1010],
[0.3,0.3821,-0.1306,-2.0305,-0.1306,0.3303,0.1306,-2.0305,0.1306,0.0943],
[0.35,-0.5810,-0.1335,-1.8905,-0.1335,0.3344,0.1335,-1.8905,0.1335,0.0936],
[0.4,-1.3119,-0.1396,-1.7568,-0.1396,0.3352,0.1396,-1.7568,0.1396,0.0969],
[0.45,-1.8796,-0.1477,-1.6339,-0.1477,0.3339,0.1477,-1.6339,0.1477,0.1030],
[0.5,-2.3275,-0.1570,-1.5236,-0.1570,0.3309,0.1570,-1.5236,0.1570,0.1115],
[0.55,-2.6844,-0.1669,-1.4264,-0.1669,0.3264,0.1669,-1.4264,0.1669,0.1218],
[0.6,-2.9708,-0.1770,-1.3418,-0.1770,0.3206,0.1770,-1.3418,0.1770,0.1339],
[0.65,-3.2020,-0.1871,-1.2691,-0.1871,0.3135,0.1871,-1.2691,0.1871,0.1475],
[0.7,-3.3893,-0.1968,-1.2073,-0.1968,0.3052,0.1968,-1.2073,0.1968,0.1626],
[0.75,-3.5417,-0.2060,-1.1552,-0.2060,0.2958,0.2060,-1.1552,0.2060,0.1791],
[0.8,-3.6660,-0.2145,-1.1117,-0.2145,0.2853,0.2145,-1.1117,0.2145,0.1968],
[0.85,-3.7675,-0.2222,-1.0758,-0.2222,0.2738,0.2222,-1.0758,0.2222,0.2157],
[0.9,-3.8505,-0.2288,-1.0466,-0.2288,0.2613,0.2288,-1.0466,0.2288,0.2356],
[0.95,-3.9183,-0.2343,-1.0233,-0.2343,0.2481,0.2343,-1.0233,0.2343,0.2564],
[1,-3.9734,-0.2385,-1.0052,-0.2385,0.2343,0.2385,-1.0052,0.2385,0.2779],
[1.05,-4.0180,-0.2414,-0.9916,-0.2414,0.2199,0.2414,-0.9916,0.2414,0.3000],
[1.1,-4.0539,-0.2430,-0.9820,-0.2430,0.2053,0.2430,-0.9820,0.2430,0.3225],
[1.15,-4.0825,-0.2431,-0.9758,-0.2431,0.1904,0.2431,-0.9758,0.2431,0.3451],
[1.2,-4.1050,-0.2418,-0.9725,-0.2418,0.1756,0.2418,-0.9725,0.2418,0.3678],
[1.25,-4.1224,-0.2392,-0.9716,-0.2392,0.1610,0.2392,-0.9716,0.2392,0.3902],
[1.3,-4.1356,-0.2353,-0.9728,-0.2353,0.1466,0.2353,-0.9728,0.2353,0.4123],
[1.35,-4.1454,-0.2301,-0.9757,-0.2301,0.1327,0.2301,-0.9757,0.2301,0.4339],
[1.4,-4.1523,-0.2239,-0.9798,-0.2239,0.1194,0.2239,-0.9798,0.2239,0.4549],
[1.45,-4.1568,-0.2167,-0.9850,-0.2167,0.1068,0.2167,-0.9850,0.2167,0.4751],
[1.5,-4.1594,-0.2086,-0.9910,-0.2086,0.0948,0.2086,-0.9910,0.2086,0.4945],
[1.55,-4.1605,-0.1998,-0.9975,-0.1998,0.0837,0.1998,-0.9975,0.1998,0.5129],
[1.6,-4.1602,-0.1905,-1.0045,-0.1905,0.0734,0.1905,-1.0045,0.1905,0.5304],
[1.65,-4.1589,-0.1807,-1.0116,-0.1807,0.0640,0.1807,-1.0116,0.1807,0.5468],
[1.7,-4.1568,-0.1707,-1.0189,-0.1707,0.0555,0.1707,-1.0189,0.1707,0.5622],
[1.75,-4.1540,-0.1605,-1.0262,-0.1605,0.0479,0.1605,-1.0262,0.1605,0.5766],
[1.8,-4.1508,-0.1503,-1.0334,-0.1503,0.0410,0.1503,-1.0334,0.1503,0.5899],
[1.85,-4.1471,-0.1403,-1.0404,-0.1403,0.0350,0.1403,-1.0404,0.1403,0.6023],
[1.9,-4.1431,-0.1305,-1.0473,-0.1305,0.0297,0.1305,-1.0473,0.1305,0.6138],
[1.95,-4.1390,-0.1210,-1.0540,-0.1210,0.0251,0.1210,-1.0540,0.1210,0.6244],
[2,-4.1347,-0.1119,-1.0605,-0.1119,0.0212,0.1119,-1.0605,0.1119,0.6342],
[2.05,-4.1303,-0.1031,-1.0667,-0.1031,0.0178,0.1031,-1.0667,0.1031,0.6432],
[2.1,-4.1258,-0.0949,-1.0727,-0.0949,0.0148,0.0949,-1.0727,0.0949,0.6516],
[2.15,-4.1214,-0.0871,-1.0785,-0.0871,0.0124,0.0871,-1.0785,0.0871,0.6594],
[2.2,-4.1169,-0.0797,-1.0840,-0.0797,0.0103,0.0797,-1.0840,0.0797,0.6666],
[2.25,-4.1125,-0.0729,-1.0893,-0.0729,0.0085,0.0729,-1.0893,0.0729,0.6733],
[2.3,-4.1082,-0.0665,-1.0944,-0.0665,0.0070,0.0665,-1.0944,0.0665,0.6796],
[2.35,-4.1040,-0.0606,-1.0993,-0.0606,0.0058,0.0606,-1.0993,0.0606,0.6854],
[2.4,-4.0998,-0.0551,-1.1040,-0.0551,0.0047,0.0551,-1.1040,0.0551,0.6909],
[2.45,-4.0957,-0.0500,-1.1085,-0.0500,0.0039,0.0500,-1.1085,0.0500,0.6961],
[2.5,-4.0918,-0.0454,-1.1128,-0.0454,0.0032,0.0454,-1.1128,0.0454,0.7010],
[2.55,-4.0879,-0.0411,-1.1170,-0.0411,0.0026,0.0411,-1.1170,0.0411,0.7056],
[2.6,-4.0841,-0.0371,-1.1210,-0.0371,0.0021,0.0371,-1.1210,0.0371,0.7099],
[2.65,-4.0805,-0.0335,-1.1248,-0.0335,0.0017,0.0335,-1.1248,0.0335,0.7141],
[2.7,-4.0769,-0.0303,-1.1285,-0.0303,0.0014,0.0303,-1.1285,0.0303,0.7181],
[2.75,-4.0735,-0.0273,-1.1321,-0.0273,0.0011,0.0273,-1.1321,0.0273,0.7218],
[2.8,-4.0701,-0.0245,-1.1356,-0.0245,0.0009,0.0245,-1.1356,0.0245,0.7254],
[2.85,-4.0669,-0.0221,-1.1389,-0.0221,0.0007,0.0221,-1.1389,0.0221,0.7289],
[2.9,-4.0638,-0.0198,-1.1421,-0.0198,0.0006,0.0198,-1.1421,0.0198,0.7322],
[2.95,-4.0607,-0.0178,-1.1452,-0.0178,0.0005,0.0178,-1.1452,0.0178,0.7354],
[3,-4.0578,-0.0159,-1.1482,-0.0159,0.0004,0.0159,-1.1482,0.0159,0.7385],
[3.05,-4.0549,-0.0142,-1.1511,-0.0142,0.0003,0.0142,-1.1511,0.0142,0.7414],
[3.1,-4.0521,-0.0127,-1.1539,-0.0127,0.0002,0.0127,-1.1539,0.0127,0.7443],
[3.15,-4.0494,-0.0114,-1.1566,-0.0114,0.0002,0.0114,-1.1566,0.0114,0.7470],
[3.2,-4.0468,-0.0101,-1.1592,-0.0101,0.0001,0.0101,-1.1592,0.0101,0.7497],
[3.25,-4.0443,-0.0090,-1.1618,-0.0090,0.0001,0.0090,-1.1618,0.0090,0.7522],
[3.3,-4.0418,-0.0081,-1.1643,-0.0081,0.0001,0.0081,-1.1643,0.0081,0.7547],
[3.35,-4.0394,-0.0072,-1.1666,-0.0072,0.0001,0.0072,-1.1666,0.0072,0.7571],
[3.4,-4.0371,-0.0064,-1.1690,-0.0064,0.0001,0.0064,-1.1690,0.0064,0.7595],
[3.45,-4.0349,-0.0056,-1.1712,-0.0056,0.0000,0.0056,-1.1712,0.0056,0.7617],
[3.5,-4.0327,-0.0050,-1.1734,-0.0050,0.0000,0.0050,-1.1734,0.0050,0.7639],
[3.55,-4.0306,-0.0044,-1.1756,-0.0044,0.0000,0.0044,-1.1756,0.0044,0.7661],
[3.6,-4.0285,-0.0039,-1.1776,-0.0039,0.0000,0.0039,-1.1776,0.0039,0.7681],
[3.65,-4.0265,-0.0035,-1.1796,-0.0035,0.0000,0.0035,-1.1796,0.0035,0.7702],
[3.7,-4.0245,-0.0030,-1.1816,-0.0030,0.0000,0.0030,-1.1816,0.0030,0.7721],
[3.75,-4.0226,-0.0027,-1.1835,-0.0027,0.0000,0.0027,-1.1835,0.0027,0.7740],
[3.8,-4.0208,-0.0024,-1.1854,-0.0024,0.0000,0.0024,-1.1854,0.0024,0.7759],
[3.85,-4.0190,-0.0021,-1.1872,-0.0021,0.0000,0.0021,-1.1872,0.0021,0.7777],
[3.9,-4.0172,-0.0018,-1.1889,-0.0018,0.0000,0.0018,-1.1889,0.0018,0.7795],
[3.95,-4.0155,-0.0016,-1.1906,-0.0016,0.0000,0.0016,-1.1906,0.0016,0.7812]]
)

#Building the Hamiltonian H[0]= Radius, H[1]=H(Radius)
H=[]
(n,m)=Hamiltonian_coef.shape
for i in range(n): #i = Radius
    h_0=1.0*np.matrix(Hamiltonian_elem[0])
    for j in range(1,m):
        h_0+= Hamiltonian_coef[i,j]*np.matrix(Hamiltonian_elem[j])
    H.append([Hamiltonian_coef[i,0],h_0])

# Simulation
tq = tqdm(desc='Minimizing...')  # Displaying progress bar
radius1 = []
E1 = []
init_param = []

for R in range(len(H)):  # We try to find the ground state eigenvalue for each radius R
    radius1.append(H[R][0])
    if (init_param == []):  #
        init_param = [2 * (np.pi) * random.random() for _ in List_Parameters]
    else:
        for i in range(len(init_param)):
            init_param[i] = VQE.get_parameters()[i]._value

    # Finding the ground state eigen value for each H(R)
    result = minimize(minimize_loss, init_param, method='Nelder-Mead')

    E1.append(result.get('fun'))
    tq.set_description('Finished')

E1_th=[]
for h in H:
    l0=np.linalg.eigvals(h[1])
    l0.sort()
    E1_th.append(min(l0))

plt.plot(100*np.array(radius1),E1_th,'orange')
plt.plot(100*np.array(radius1),E1,'x')
plt.ylabel('Energy (MJ/mol)')
plt.xlabel('Atomic separation R (pm)')
plt.legend(['Theoretical eigenvalues', 'Eigenvalues computed with Perceval'])
plt.show()

plt.plot(100*np.array(radius1),E1_th,'orange')
plt.plot(100*np.array(radius1),E1,'x')
plt.axis([50,250,-5.8,-5.5])
plt.ylabel('Energy (MJ/mol)')
plt.xlabel('Atomic separation R (pm)')

plt.legend(['Theoretical eigenvalues', 'Eigenvalues computed with Perceval'])

plt.show()

min_value=min(E1)
min_index = E1.index(min_value)
print('The minimum energy is E_g('+str(radius1[min_index])+')='+str(E1[min_index])+' MJ/mol and is attained for R_min ='+str(radius1[min_index])+' pm')
\end{lstlisting}

\subsection{Quantum Machine Learning - Differential Equation Solving}

\begin{lstlisting}[language=python,style=mypython,caption=Example code - \ref{sec:qml} Quantum Machine Learning,captionpos=b,label=code:appQML]
import perceval as pcvl
import numpy as np
from math import comb
from scipy.optimize import minimize
import time
import matplotlib.pyplot as plt
import matplotlib as mpl
import tqdm as tqdm

nphotons = 4

# Differential equation parameters
lambd = 8
kappa = 0.1

def F(u_prime, u, x):       # DE, works with numpy arrays
    return u_prime + lambd * u * (kappa + np.tan(lambd * x))

# Boundary condition (f(x_0)=f_0)
x_0 = 0
f_0 = 1

# Modeling parameters
n_grid = 50    # number of grid points of the discretized differential equation
range_min = 0  # minimum of the interval on which we wish to approximate our function
range_max = 1  # maximum of the interval on which we wish to approximate our function
X = np.linspace(range_min, range_max-range_min, n_grid)  # Optimisation grid

# Differential equation's exact solution - for comparison
def u(x):
    return np.exp(- kappa*lambd*x)*np.cos(lambd*x)

# Parameters of the quantum machine learning procedure
N = nphotons              # Number of photons
m = nphotons              # Number of modes
eta = 5                   # weight granted to the initial condition
a = 200                   # Approximate boundaries of the interval that the image of the trial function can cover
fock_dim = comb(N + m - 1, N)
# lambda coefficients for all the possible outputs
lambda_random = 2 * a * np.random.rand(fock_dim) - a
# dx serves for the numerical differentiation of f
dx = (range_max-range_min) / (n_grid - 1)

# Input state with N photons and m modes
input_state = pcvl.BasicState([1]*N+[0]*(m-N))

"Haar unitary parameters"
# number of parameters used for the two universal interferometers (2*m**2 per interferometer)
parameters = np.random.normal(size=4*m**2)

px = pcvl.P("px")
c = pcvl.Unitary(pcvl.Matrix.random_unitary(m, parameters[:2 * m ** 2]), name="W1")\
     // (0, pcvl.PS(px))\
     // pcvl.Unitary(pcvl.Matrix.random_unitary(m, parameters[2 * m ** 2:]), name="W2")

simulator_backend = pcvl.BackendFactory().get_backend("SLOS")
s1 = simulator_backend(pcvl.Matrix.random_unitary(m))
s1.compile(input_state)
pcvl.pdisplay(c)

def computation(params):
    global current_loss
    global computation_count
    "compute the loss function of a given differential equation in order for it to be optimized"
    computation_count += 1
    f_theta_0 = 0  # boundary condition
    coefs = lambda_random  # coefficients of the M observable
    # initial condition with the two universal interferometers and the phase shift in the middle
    U_1 = pcvl.Matrix.random_unitary(m, params[:2 * m ** 2])
    U_2 = pcvl.Matrix.random_unitary(m, params[2 * m ** 2:])

    px = pcvl.P("x")
    c = pcvl.Unitary(U_2) // (0, pcvl.PS(px)) // pcvl.Unitary(U_1)

    px.set_value(np.pi * x_0)
    U = c.compute_unitary(use_symbolic=False)
    s1.U = U
    f_theta_0 = np.sum(np.multiply(s1.all_prob(input_state), coefs))

    # boundary condition given a weight eta
    loss = eta * (f_theta_0 - f_0) ** 2 * len(X)

    # Y[0] is before the domain we are interested in (used for differentiation), x_0 is at Y[1]
    Y = np.zeros(n_grid + 2)

    # x_0 is at the beginning of the domain, already calculated
    Y[1] = f_theta_0

    px.set_value(np.pi * (range_min - dx))
    s1.U = c.compute_unitary(use_symbolic=False)
    Y[0] = np.sum(np.multiply(s1.all_prob(input_state), coefs))

    for i in range(1, n_grid):
        x = X[i]
        px.set_value(np.pi * x)
        s1.U = c.compute_unitary(use_symbolic=False)
        Y[i + 1] = np.sum(np.multiply(s1.all_prob(input_state), coefs))

    px.set_value(np.pi * (range_max + dx))
    s1.U = c.compute_unitary(use_symbolic=False)
    Y[n_grid + 1] = np.sum(np.multiply(s1.all_prob(input_state), coefs))

    # Differentiation
    Y_prime = (Y[2:] - Y[:-2])/(2*dx)

    loss += np.sum((F(Y_prime, Y[1:-1], X))**2)

    current_loss = loss / len(X)
    return current_loss

def callbackF(parameters):
    """callback function called by scipy.optimize.minimize allowing to monitor progress"""
    global current_loss
    global computation_count
    global loss_evolution
    global start_time
    now = time.time()
    pbar.set_description("M= %d Loss: %0.5f #computations: %d elapsed: %0.5f" %
                         (m, current_loss, computation_count, now-start_time))
    pbar.update(1)
    loss_evolution.append((current_loss, now-start_time))
    computation_count = 0
    start_time = now

computation_count = 0
current_loss = 0
start_time = time.time()
loss_evolution = []

pbar = tqdm.tqdm()
res = minimize(computation, parameters, callback=callbackF, method='BFGS', options={'gtol': 1E-2})
print("Unitary parameters", res.x)

def plot_solution(m, N, X, optim_params, lambda_random):
    Y = []
    U_1 = pcvl.Matrix.random_unitary(m, optim_params[:2 * m ** 2])
    U_2 = pcvl.Matrix.random_unitary(m, optim_params[2 * m ** 2:])
    px = pcvl.P("x")
    c = pcvl.Unitary(U_2) // (0, pcvl.PS(px)) // pcvl.Unitary(U_1)

    for x in X:
        px.set_value(np.pi * x)
        U = c.compute_unitary(use_symbolic=False)
        s1.U = U
        f_theta = np.sum(np.multiply(s1.all_prob(input_state), lambda_random))
        Y.append(f_theta)
    exact = u(X)
    plt.plot(X, Y, label="Approximation with {} photons".format(N))

X = np.linspace(range_min, range_max, 200)

# Change the plot size
default_figsize = mpl.rcParamsDefault['figure.figsize']
mpl.rcParams['figure.figsize'] = [2 * value for value in default_figsize]

plot_solution(m, N, X, res.x, lambda_random)

plt.plot(X, u(X), 'r', label='Analytical solution')
plt.legend()
plt.show()

plt.plot([v[0] for v in loss_evolution])
plt.yscale("log")
plt.xlabel("Number of epochs")
plt.ylabel("Loss function value")
\end{lstlisting}